# Quantifying inconsistency in one-stage individual participant data meta-analyses of treatment-covariate interactions: a simulation study

Myra B. McGuinness*,[1,2,3] Joanne E. McKenzie,[4] Andrew Forbes,[5] Flora Hui,[2,3] Keith R. Martin,[2,3] Robert J. Casson,[6] Amalia Karahalios[1]

1 Centre for Epidemiology and Biostatistics, Melbourne School of Population and Global Health, University of Melbourne, Melbourne, Australia

2 Centre for Eye Research Australia, Royal Victorian Eye and Ear Hospital, Melbourne, Australia

3 Department of Surgery (Ophthalmology), Melbourne Medical School, University of Melbourne, Melbourne, Australia

4 Methods in Evidence Synthesis Unit, School of Public Health and Preventive Medicine, Monash University.

5 Biostatistics Unit, School of Public Health and Preventive Medicine, Monash University, Melbourne Australia

6 Ophthalmic Research Laboratories, Discipline of Ophthalmology and Visual Sciences, University of Adelaide, Adelaide, Australia

* Corresponding author: Myra B. McGuinness myra.mcguinness@unimelb.edu.au

**Declaration of conflicting interests**

The Authors declare that there is no conflict of interest.

**Funding**

This research received no specific grant from any funding agency in the public, commercial, or not-for-profit sectors. The Centre for Eye Research Australia receives operational infrastructure support from the Victorian State government.

**Keywords**

Meta-analysis

Heterogeneity

Individual participant data

**Abstract**

It is recommended that measures of between-study effect heterogeneity be reported when conducting individual-participant data meta-analyses (IPD-MA). Methods exist to quantify inconsistency between trials via $I^2$ (the percentage of variation in the treatment effect due to between-study heterogeneity) when conducting two-stage IPD-MA, and when conducting one-stage IPD-MA with approximately equal numbers of treatment and control group participants. We extend formulae to estimate $I^2$ when investigating treatment-covariate interactions with unequal numbers of participants across subgroups and/or continuous covariates. A simulation study was conducted to assess the agreement in values of $I^2$ between those derived from two-stage models using traditional methods and those derived from equivalent one-stage models. Fourteen scenarios differed by the magnitude of between-trial heterogeneity, the number of trials, and the average number of participants in each trial. Bias and precision of $\hat{I}^2$ were similar between the one- and two-stage models. The mean difference in $\hat{I}^2$ between equivalent models ranged between -1.0 and 0.0 percentage points across scenarios. However, disparities were larger in simulated datasets with smaller samples sizes with up to 19.4 percentage points difference between models. Thus, the estimates of $I^2$ derived from these extended methods can be interpreted similarly to those from existing formulae for two-stage models.

# 1   Introduction

## 1.1   Background

The number of individual participant data meta-analyses (IPD-MA) published each year continues to grow,[1] with further growth expected as public research funders encourage data sharing. The use of individual-level data allows standardisation of variable definitions, analysis populations, and modelling approaches across trials. Despite this, observed treatment effects are still likely to differ between trials.[2] Thus, current reporting guidelines for IPD-MA recommend that measures of between-trial effect heterogeneity be provided along with pooled estimates of treatment effect.[3]

The $I^2$ statistic is commonly reported in meta-analyses as a measure of inconsistency between trials, and represents the percentage of variation in the treatment effect due to between-study heterogeneity.[4] Methods exists to estimate $I^2$ when conducting two-stage IPD-MA (i.e., when estimates are derived from each trial separately prior to pooling), and when conducting one-stage IPD-MA with approximately equal numbers of treatment and control group participants.[5] However, many IPD-MA are intended to investigate treatment-covariate interactions with unequal number of participants in each covariate subgroup, and some interactions may involve continuous covariates.[6-8]

Individual trials should be powered to detect a clinically meaningful difference in the primary outcome between treatment groups. However, they often lack the sample size required to detect differences in treatment effect between key demographic groups. Hence, prospective IPD-MA may be planned to investigate which subgroups of participants benefit most from the treatment.[6, 8] In clinical trial research this difference in treatment effect is often referred to as treatment effect heterogeneity. In this paper, we use the term treatment-covariate interaction instead to avoid confusion with between-trial heterogeneity.

Below we describe steps for deriving $I^2$ from a one-stage random-effects IPD-MA with a continuous outcome intended to estimate the difference in treatment effect between participant subgroups. We then present a simulation study to assess how well estimates of $I^2$ derived from these methods agree with those calculated using traditional methods designed for aggregate-data (or two-stage) meta-analyses. Specifically, we aimed to estimate the mean difference in values of $I^2$ derived from one- and two-stage IPD-MA approaches under different conditions. Secondary objectives included quantifying the bias, empirical standard error, and mean square error for each of the average within-trial effect variance, between-trial effect variance, and $I^2$ estimates from each model.

**1.2   Motivating example**

Four ongoing parallel-group trials aim to investigate the effect of an oral vitamin B3 supplement (nicotinamide) on the rate of visual field loss among people with glaucoma in Australia (Australian New Zealand Clinical Trials Registry ACTRN12622000414718), Sweden (clinicaltrials.gov NCT05275738), the UK (clinicaltrials.gov NCT05405868), and Singapore (Health Sciences Authority CTC2400074). The primary outcome variable, visual field mean deviation, represents the average sensitivity to light in an area around the central vision. As previously recommended for trials of glaucoma interventions that aim to slow visual field loss,[9,10] the rate of change over repeated visits (measured in decibels/year) will be compared between active and placebo treatment groups for the primary analysis of each trial. This will be conducted via mixed effects linear regression with an interaction term between treatment group and time, and will include data from all study visits over the two-year study period.[11] Although change in visual field sensitivity is expected to be curvilinear over a person's lifetime, we expect that any change in mean deviation will be approximately linear over the two-year study period.[12]

In each trial, random treatment allocation has been stratified by glaucoma subtype (primary open angle or pseudoexfoliative glaucoma) as investigators believe it may be a treatment-effect modifier. A prospective IPD-MA is planned to investigate treatment-covariate interactions for this and other baseline characteristics.

## 2   Deriving $I^2$ from one-stage random-effects IPD-MA

### 2.1   $I^2$ in aggregate data meta-analyses

In random-effects meta-analyses aiming to estimate the pooled treatment-covariate interaction effect $\hat{\theta}$, the total variance of the effect can be decomposed into that due to within-trial variance, $\bar{\sigma}_\theta^2$, and that due to between-trial variance, $\tau^2$. $I^2$ represents the percentage of total variability in the treatment effect due to between-trial heterogeneity:[4]

$$I^2 = 100 \times \frac{\hat{\tau}^2}{\hat{\tau}^2 + \bar{\sigma}_\theta^2} \quad (1)$$

In aggregate data meta-analyses and two-stage IPD-MA, the variance of the estimated treatment-covariate interaction within trial $j$, $\hat{\sigma}_{\theta j}^2$, can be used to calculate the fixed-effect inverse variance weight for that trial:

$$w_{Fj} = \frac{1}{\hat{\sigma}_{\theta j}^2} \quad (2)$$

These weights can be used to estimate the average within-trial variance of the treatment effect across $K$ trials ($\bar{\sigma}_\theta^2$) which is required for Equation 1:[4]

$$\bar{\sigma}_\theta^2 = \frac{K-1}{\sum_{j=1}^{K} w_{Fj} - \left(\frac{\sum_{j=1}^{K} w_{Fj}^2}{\sum_{j=1}^{K} w_{Fj}}\right)} \quad (3)$$

### 2.2 Approximating within-trial interaction effect variance in one-stage IPD-MA

One-stage IPD-MA models do not produce separate estimates of effect variance ($\hat{\sigma}_{\theta j}^2$) for each trial. Formulae for approximating $\hat{\sigma}_{\theta j}^2$ have previously been provided for studies with a single outcome visit and approximately equal numbers of participants in the active and control treatment groups.[6, 7] If the treatment group and covariate variables are independent (as expected following randomisation) the following equation provides an alternative that can be used for studies with repeated measures and/or continuous covariates, and simplifies to the existing formula for studies with a single outcome visit:

$$\hat{\sigma}_{\theta j}^2 = \frac{\hat{\sigma}_{ej}^2}{n_{(obs)j} \prod_{p=1}^{P} var(X_{jp})} \quad (4)$$

where $\hat{\sigma}_{ej}^2$ is the residual error variance for trial $j$ estimated via the IPD-MA model,

$n_{(obs)j}$ is the number of observations in the contrast of interest,

$P$ is the number of variables in the contrast of interest, and

the variance of each of those variables is $var(X_{jp}) = \frac{\sum_{i=1}^{no_j}(x_{ijp} - \bar{x}_{jp})^2}{n_{(obs)j}}$.

The derivation of $\hat{\sigma}_{\theta j}^2$ from a one-stage model with a treatment-covariate interaction (Equation 4) is provided in Supplemental Material A. The value of $\hat{\sigma}_{\theta j}^2$ from Equation 4 can be used to derive inverse variance weights ($w_{Fj}$, Equation 2) and average within-trial variance ($\bar{\sigma}_\theta^2$, Equation 3) to allow estimation of $I^2$ (Equation 1)

When a meta-analysis aims to quantify the magnitude of a treatment-covariate interaction between a binary treatment group variable $A$ and binary disease subgroup variable $Z$ at a single outcome visit, the analysis model will often include fixed effect terms for each of these variables in addition to the interaction term ($A \times Z$). Here the contrast of interest is the difference in treatment effect between disease subtypes and $P = 2$ (so that $var(X_{j1})$ is the variance of the treatment variable, $A$, and $var(X_{j2})$ is variance of the subgroup variable, $Z$). In a setting in which each participant contributes one observation to the analysis, $n_{(obs)j}$ is equal to the number of participants in trial $j$, $n_j$.

For analyses of repeated measures in which the contrast of interest relates to rate of change over time, $n_{(obs)j} = \sum_{i=1}^{n_j} G_{ij}$ where $G_{ij}$ is the number of visits attended by participant $i$ in trial $j$, and time ($T$) is included in the $P$ interaction variables. Note that for models in which a main effect coefficient is constrained to zero (e.g., the treatment group coefficient in a repeated measures model when no difference is expected at baseline due to randomisation), the contrast of interest, number of covariates ($P$) and number of observations ($n_{(obs)j}$) will change.

Some one-stage models may fail to converge when all parameters are included as specified in the two-stage model, particularly when the number of trials is large. When this is the case, a more parsimonious one-stage model may be fit in which the residual error variance is assumed to be common across all trials ($\hat{\sigma}_e^2$) rather than estimated separately for each trial ($\hat{\sigma}_{ej}^2$).[13] In that case, $\hat{\sigma}_e^2$ can be substituted for $\hat{\sigma}_{ej}^2$ into Equation 4.

## 3   Simulation study methods

A prospectively registered protocol for this simulation study is available online.[14] Data generation and analysis was conducted using Stata/MP 18 (StataCorp, College Station, TX) with the 64-bit Mersenne twister for random number generation.[15] The two-stage models were fit via the `ipdmetan` command,[16] as seen in the example statistical computing code provided in Supplemental Material B.

### 3.1   Target and estimands

The primary target in this simulation study is $I^2$, the percentage of variation in the treatment-covariate interaction due to between-trial heterogeneity. Performance was investigated separately for two clinical research questions (Estimands 1 & 2).

The effect of interest for Estimand 1 is the treatment-subgroup interaction at the final study visit, i.e., the mean difference between disease subtypes (primary open angle, $Z = 0$, vs pseudoexfoliative glaucoma, $Z = 1$) in the treatment effect on visual field mean deviation two years from randomisation ($T = 2$). For Estimand 1, the contrast of interest is:

$$\theta = E(Y_{A=1} - Y_{A=0}|T = 2, Z = 1) - E(Y_{A=1} - Y_{A=0}|T = 2, Z = 0) \quad (5)$$

where $Y_{A=a}$ represents the outcome for treatment group $A = a$ ($a = 0,1$).

To align with the motivating example which aims to assess the rate of change in the outcome variable over time, a second model was fit for a longitudinal dataset with repeated outcome measures. The effect of interest for Estimand 2 is the treatment-time-subgroup interaction, i.e., the mean difference between disease subtypes in the treatment effect on rate of change in visual field mean deviation over the two-year study period. For Estimand 2, the contrast of interest is:

$$\theta = E(Rate_{A=1} - Rate_{A=0}|Z=1) - E(Rate_{A=1} - Rate_{A=0}|Z=0) \tag{6}$$

where $Rate_{A=a}$ is the average change in the outcome per year for treatment group $A=a$.

### 3.2   Data generating mechanisms

For each parallel-group trial $j$ ($j = 1, ..., K$), glaucoma subtype $Z$ ($Z = 0$ for primary open angle, $Z = 1$ for pseudoexfoliative) was randomly generated within the $n_{ja}$ participants in the active treatment group ($A = 1$), with $\Pr(Z = 1) = 0.375$ across all trials (so the proportion with $Z = 1$ differed between trials and replicated datasets). $n_{ja}$ matched participants were generated in the placebo treatment group ($A = 0$) so that the treatment was allocated 1:1 (stratified by disease subtype) and the total number of participants in trial $j$, $n_j$, was equal to $2n_{ja}$. Within each replicated dataset, sample size in each treatment arm was randomly generated with a mean of $\bar{n}_{ja}$ and a standard deviation of 10 so that sample sizes differed between trials and replications. $T$ represents time from randomisation in years with $G = 5$ study visits equally spaced across and within trials ($t = 0(0.5)2$).

The clinical outcome variable (visual field mean deviation) was generated for a single eye of participant $i$ at time $t$ in trial $j$ using the following formula:

$$y_{itj} = a_{ij}z_{ij}t(\beta_1 + u_{1j}) + \beta_2 t + \beta_3 a_{ij}t + u_{2ij} + e_{itj} \tag{7}$$

where:

$\beta_1$ is the difference in treatment effect (active - placebo) on rate of change per year between disease subtypes for the underlying population ($\beta_1 = 0.12$),

$\beta_2$ is the change in outcome per one-year change in $T$ among the control group with no pseudoexfoliative glaucoma ($\beta_2 = 0.4$),

$\beta_3$ is the difference in change in outcome per year between intervention groups among those without pseudoexfoliative glaucoma ($\beta_3 = 0.12$),

$u_{1j}$ is the random trial-specific treatment-time-subgroup interaction effect with distribution $\sim^{\text{iid}} N[0, \sigma_{u1}^2]$,

$u_{2ij}$ is a random intercept terms for participant $i$ in trial $j$ with distribution $\sim^{\text{iid}} N(0, \sigma_{u2}^2 = 4)$, and $e_{itj}$ is the random model error/residual with distribution $\sim^{\text{iid}} N(0, \sigma_e^2 = 1)$.

Note that the analysis models (specified in Section 3.4) had additional fixed effect parameters and their coefficient values are expected to equal zero in an unbiased model.

### 3.3 Scenario factor values and replications

Conditions were varied in a partially factorial manner so that 14 scenarios were produced according to number of trials ($K = 4, 16, 30$), average sample size in each trial ($\bar{n}_j = 90, 400$), and treatment-effect heterogeneity for the treatment-time-subgroup effect ($\sigma_{u1}^2 = 0.00, 0.01, 0.05$). As pre-specified,[14] Estimand 2 was only assessed in scenarios with <30 trials due to the high computational demand associated with the repeated measures model.

The true population values of $\tau^2$, $\bar{\sigma}_\theta^2$, and $I^2$ were algebraically derived and are presented in Supplemental Tables 3-5.

Datasets were generated 1000 times per scenario. This number of replications was chosen to provide sufficient precision for clinical interpretation of the agreement in $I^2$ between models, given the Monte Carlo standard error observed in pilot results.[14]

### 3.4 Analysis of each simulated dataset

Three random-effects models were fit for each estimand: a two-stage model (M2), a full one-stage model with independent residual error variances estimated separately for each trial (M1s, equivalent to the two-stage model), and a parsimonious one-stage model in which the residual error variance was assumed to be common across all trials (M1c).

#### 3.4.1 Two-stage random-effects IPD-MA

Two-stage IPD-MA involves deriving estimates from each trial separately prior to pooling those estimates using traditional (aggregate-data) meta-analysis methods.

One observation per participant was included in the analysis for Estimand 1 (difference between subgroups in treatment effect at two years). In the first stage, linear regression was used to estimate the magnitude of the treatment-subgroup interaction effect separately for each trial. The linear regression model was specified with fixed effect parameters for the interaction between treatment and subgroup ($A \times Z$, the clinical effect parameter of interest, $\hat{\beta}_{1j} = \hat{\theta}_j$), the main effect for each of these variables separately ($A$ and $Z$), and the baseline value of the outcome variable ($Y0$):

$$y_{ij(t=2)} = \hat{\beta}_{0j} + \hat{\beta}_{1j}a_{ij}z_{ij} + \hat{\beta}_{2j}a_{ij} + \hat{\beta}_{3j}z_{ij} + \hat{\beta}_{4j}y0_{ij} + e_{ij} \qquad (8)$$

Five observations per participant (one from each study visit) were included in the analysis for Estimand 2 (difference between subgroups in treatment effect on rate of change over time). In the first stage, mixed-effects restricted maximum likelihood (REML) linear regression was used to estimate the treatment-time-subgroup interaction effect separately for each trial. Fixed effect parameters were included for the interaction between treatment, time, and subgroup ($A \times Z \times T$, the clinical effect parameter of interest), pairwise interactions for each of these variables ($A \times Z$, $A \times T$, and $Z \times T$), and the main effect for each of these variables separately ($A$, $Z$, and $T$). Time was included as a continuous covariate. A random intercept parameter with identity covariance structure was used to account for the non-independence of repeated measures for each participant ($u_{ij} \sim N(0, \sigma^2_{uij})$), and independent residual model errors were specified ($e_{itj} \sim N(0, \sigma^2_{ej})$):

$$y_{itj} = (\hat{\beta}_{0j} + u_{ij}) + \hat{\beta}_{1j}a_{ij}z_{ij}t + \hat{\beta}_{2j}a_{ij} + \hat{\beta}_{3j}z_{ij} + \hat{\beta}_{4j}t + \hat{\beta}_{5j}a_{ij}z_{ij} + \hat{\beta}_{6j}a_{ij}t + \hat{\beta}_{7j}z_{ij}t + e_{itj} \quad (9)$$

For each estimand, between-trial heterogeneity $\hat{\tau}^2$ was estimated iteratively via REML in the second stage of the IPD-MA and profile likelihood 95% CIs for $\hat{\tau}^2$ were obtained as executed by the `ipdmetan` command.[16, 17] $\hat{I}^2$ was calculated using the average within-trial fixed-effect variance (Equation 3) rather than using output from the `ipdmetan` command which is derived using the $Q$ statistic.[16]

In the second stage of a two-stage random-effects IPD-MA, the pooled estimate for the treatment-covariate interaction $\hat{\theta}$ is derived using random-effects inverse variance trial weights ($1/[\hat{\sigma}^2_{\theta j} + \hat{\tau}^2]$, although random effects weights were not derived or extracted as part of this simulation study).

### 3.4.2   One-stage random-effects IPD-MA

In a one-stage IPD-MA, data from all trials are analysed using a single model. Mixed-effects REML models were used for the one-stage models in this simulation study. Each model included a fixed effect parameter for the primary treatment-covariate interaction of interest to each estimand, representing the pooled interaction effect across all trials ($\hat{\beta}_1 = \hat{\theta}$). Trial-specific variables were included as model covariates for each of the other fixed effect parameters specified above ($\hat{\beta}_{2j}$-$\hat{\beta}_{4j}$ from Equation 8 for Estimand 1, $\hat{\beta}_{2j}$-$\hat{\beta}_{7j}$ from Equation 9 for Estimand 2). For each trial, the trial-specific variable took on the value of the covariate $(X_j = X)$ when $J = j$ and zero otherwise. An indicator variable for each trial ($Trial_j = 1$ when $J = j$ and zero otherwise) was also included to produce trial-specific intercept parameters ($\hat{\beta}_{0j}$).

A random slope parameter was included for the primary interaction variable to estimate the between-trial variance (i.e., effect heterogeneity, $\tau^2$) but no random trial-level intercept parameter was fitted. For Estimand 2 (treatment-subgroup-time effect), a random participant-level intercept parameter with identity covariance structure was used to account for the non-independence of repeated measures. For the full one-stage model (M1s) separate independent residual model error variance terms were estimated for each trial ($\hat{\sigma}_{ej}^2$), while a common independent residual model error variance term ($\hat{\sigma}_e^2$) was used for the parsimonious one-stage model (M1c).

### 3.5  Performance measures

Formulae for each of the performance measures are presented in the protocol.[14]

Bias was calculated as the difference between the observed and true population value for each of $\tau^2$, $\bar{\sigma}_\theta^2$, and $I^2$ in each replicated dataset. The average bias, Monte Carlo standard error (MCSE) for the average bias, empirical standard error (a measure of the precision/efficiency of the estimator), and the mean square error (a measure of accuracy which combines information about bias and empirical standard error) were calculated for each of these parameters within each scenario.

Coverage for $\tau^2$ represents the proportion of replicated datasets where the true value of $\tau^2$ was within the 95% confidence interval among models with non-missing upper and lower limit values.

The primary parameter of interest in this simulation study is $I^2$. The estimate of $I^2$ derived from the two-stage model using traditional methods designed for aggregate-data meta-analyses was considered the reference value. The difference between this value and the estimate of $I^2$ from each one-stage model was calculated ($I_\Delta^2$) within each replicated dataset to assess how well estimates from the newly described methods for one-stage IPD-MA agree with those from the traditional methods. Those differences were averaged across all datasets within each scenario and presented with the MCSE of the mean difference. Given the distribution of $I_\Delta^2$ values was left skewed for M1s (the full one-stage model), median difference and interquartile range were also derived within each scenario (rather than deriving limits of agreement from the standard deviation of the differences).

## 4  Simulation study results

### 4.1  Scenario characteristics

Cleaned data extracted from each simulated dataset can be accessed online (https://doi.org/10.17605/OSF.IO/W3SQE).

No error codes were recorded for any replicated dataset and most one-stage models converged within the 40-iteration limit (M1c 98.8%, M1s 98.8%, see Supplemental Table 1). The full one-stage model (M1s) took considerably longer to run than the parsimonious one-stage model (M1c) and the two-stage model (M2), particularly as the number of trials and participants increased (see Supplemental Table 2).

### 4.2  Bias, precision, accuracy and coverage

Biased estimates of $I^2$ were observed across all scenarios and models (average bias ranged from -20.8 to +14.8 percentage points, see Supplemental Table 3). $I^2$ tended to be overestimated in scenarios with no or low true heterogeneity and underestimated in scenarios with high true heterogeneity. In general, bias and empirical standard error decreased (i.e., accuracy increased) for $I^2$ as the number of trials increased.

Bias was also observed for estimates of $\tau^2$ and average within-trial variance ($\bar{\sigma}_\theta^2$), with relatively more bias observed for $\tau^2$ than average within-trial variance (see Supplemental Tables 4 and 5). For both parameters, bias decreased and efficiency improved as the total number of participants across all trials increased. In general, $\hat{\tau}^2$ became less biased as the true value of $\tau^2$ increased.

Bias, precision, and accuracy for $\tau^2$, $\bar{\sigma}_\theta^2$, and $I^2$ were very similar between the one- and two-stage models within each scenario (see Supplemental Figures 1-3).

Confidence intervals around $\hat{\tau}^2$ were not produced for a quarter of the one-stage models (25.4% M1s, 25.1% M1c), with greater proportions of missing lower and/or upper limits in scenarios with no true between-trial heterogeneity (see Supplemental Table 6). In general, among datasets with non-missing confidence intervals for $\hat{\tau}^2$, coverage was closer to nominal levels for scenarios with a higher number of observations and those with high levels of true heterogeneity (see Supplemental Figure 4).

### 4.3  Agreement in $I^2$ between one- and two-stage models

On average, agreement in $\hat{I}^2$ was high between the two-stage model and each one-stage model (mean difference in each scenario ranged between -1.3 and 0.0 percentage points, see

Supplemental Table 7). The lowest levels of agreement were observed for scenarios with small sample sizes in combination with a higher number of trials (see Figure 1).

Agreement with the two-stage model was more consistent for the full one-stage model (M1s, MCSE for $I_\Delta^2$ ranging from 0.00 to 0.05, $I_\Delta^2$ in individual replicated datasets ranging from -19.4 to +3.5 percentage points) than for the parsimonious one-stage model (M1c, MCSE ranging from 0.01 to 0.15, $I_\Delta^2$ in individual datasets ranging from -43.1 to +43.3 percentage points, see Supplemental Figures 5 and 6).

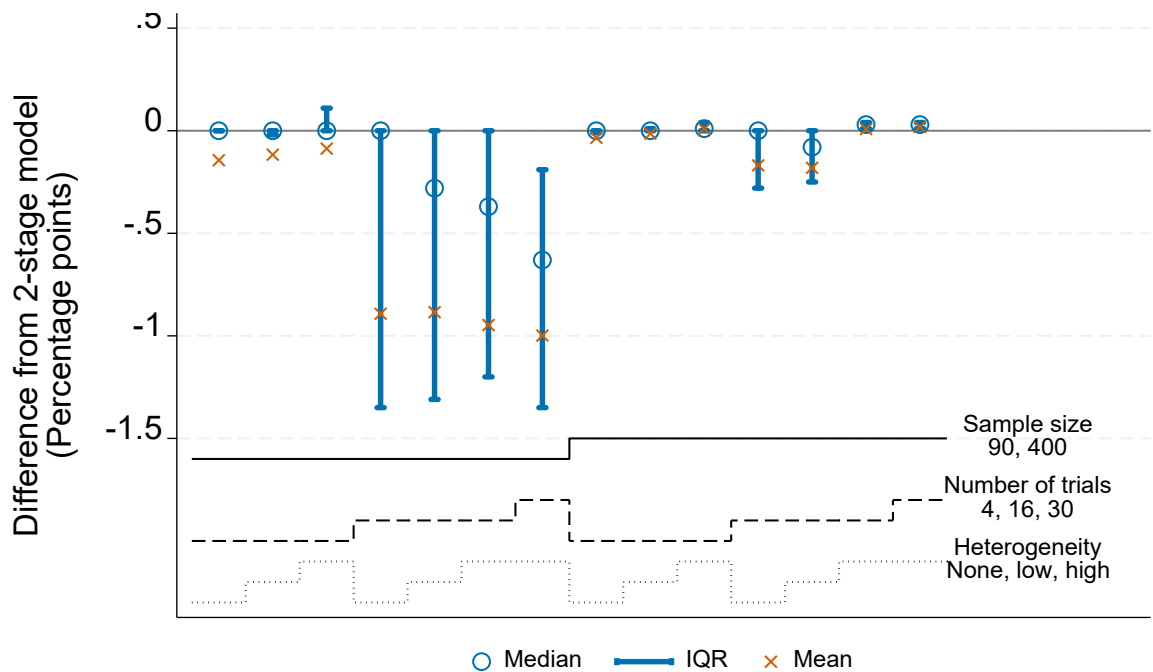

Figure 1: Difference in $\hat{I}^2$ for Estimand 1 (treatment-covariate interaction at two years) between the two-stage model and the full one-stage model with trial-level residual error variance (M1s). Negative values indicate the estimate from the 1-stage model was less than that from the 2-stage model. Solid, dashed and dotted lines indicate scenario factors in each of the 14 scenarios.

## 5   Discussion

In this paper, we have shown that between-trial inconsistency can be quantified when conducting one-stage random-effects IPD-MA of treatment-covariate interactions and continuous outcomes. In the simulation study, only minimal differences were observed in the estimates of $I^2$ derived from one- and two-stage IPD-MA when equivalent models were fit. The agreement with $\hat{I}^2$ derived from the full two-stage model remained high even when a more parsimonious one-stage model was fit.

$I^2$ is known to be unstable when there is a small number of trials.[18] In this study, estimates of $I^2$ were biased across scenarios and models. This is related to floor effects as $\hat{\tau}^2$ is truncated at zero. Thus, $I^2$ cannot be lower than 0% and the upper limit of 100% can never be achieved.[19] However, the level of bias and precision in $I^2$ did not differ considerably between the one- and two-stage models in this simulation study.

In applied settings, researchers may prefer to use a two-stage approach given the additional complexity associated with correctly specifying one-stage IPD-MA.[20] However, one-stage approaches may be favoured when there are trials with small sample sizes and the normality of error variance from each trial cannot be assumed, or when the effect of additional covariates is assumed to be common across studies.[21] Hence, the ability to estimate $I^2$ for one-stage models will be beneficial in these cases. One-stage models are also recommended for MA of binary and time-to-event outcomes when the event of interest is rare.[22] However, the methods presented in this paper only apply to continuous outcomes.

Assessment of inconsistency contributes to the interpretation of findings along with the magnitude and precision of the pooled effect size, the risk of bias, and the direction of effect within each trial.[23] Upon observing large degrees of heterogeneity, consideration must be given to the underlying causes of difference in treatment effect and whether it is sensible to pool results. Given $\hat{\tau}^2$ is presented on the measurement scale of the outcome variable, it has been recommended that it always be reported along with $\hat{I}^2$ to facilitate assessment of the clinical relevance of any heterogeneity detected while conducting random-effects MA.[4, 24] CI around $\hat{\tau}^2$ and $\hat{I}^2$ (uncertainty intervals) are likely to be informative, given the upper limit can indicate substantial heterogeneity, even with low point estimates.[25] However, we found that 95% confidence limits for $\hat{\tau}^2$ were not produced for the one-stage models in a considerable number of simulated datasets.

### 5.1   Comparison to previous findings

Hemmings et al. have shown that $I^2$ could be approximated in a one-stage IPD-MA when sample sizes are approximately even between the treatment and control arms.[5] Their approach produced estimates of $I^2$ that were highly correlated between the one- and two-stage approaches. However, given their approach was based on a one-stage model with a common residual error variance (more parsimonious than the two-stage model), the agreement between one- and two-stage models was not as high as we have demonstrated using a full one-stage model with residual error variance estimated separately for each trial. Like the current study, they showed that agreement between one- and two-stage estimates of $I^2$ increased as the total number of participants increased, and that estimates of $I^2$ were more biased for lower levels of heterogeneity.

Alternatives to $I^2$ for one-stage IPD-MA have been proposed.[26] These include a ratio of pooled treatment effect variance between random- and fixed-effect models ($R^2$),[4] and a ratio containing the modified $H^2$ statistic.[19, 27] These values have a monotonic relationship with $I^2$ as traditionally estimated, but cannot be interpreted in an identical manner.[19] This makes comparisons with other MA which have used the traditional definition of $I^2$ more complex. In contrast, the values of $I^2$ derived using the methods in the current paper can be directly compared with those derived from aggregate-data MA. Despite this, values of $I^2$ produced when investigating different research questions should be compared with caution, as $I^2$ is influenced by within-trial precision,[24] a parameter likely to differ between settings.

The methods in the current study were applied to a random-effects IPD-MA, with trial weights derived for continuous outcome variables. Alternative approaches have been proposed for producing study weights and estimates of $I^2$ for IPD-MA with binary and time-to-event outcomes.[28]

Past approaches to producing forest plots from one-stage meta-analyses have used the number of participants in each trial to derive study weights.[27] Use of the formulae presented in the current study would produce weights that more accurately reflect variance of the treatment effect.

### 5.2  Strengths and limitations

The strengths of the simulation study include the prospectively published protocol.[14] The methods were applied to a pooled estimate derived from a three-way interaction term which included both a binary and continuous covariate. We assumed a linear relationship between time and the outcome, further work is required to examine non-linear treatment-covariate interactions.[8]

We investigated the performance of the study target under several conditions, but other potentially important factors such as the magnitude of residual error variance were not varied across scenarios. In addition, keeping the total number of participants constant across scenarios would have facilitated assessment of the relative impact of the varying number of trials versus the number of participants within trials.

Estimation of within-trial variance via the methods presented in this paper for one-stage IPD-MA does require an additional step to estimate the variance of the treatment and covariate variables which is not required when conducting two-stage IPD-MA. However, this step is easily implemented in standard statistical software. The derivation of the formula used to estimate within-trial variance only holds when the treatment variable is independent of the covariate. This is likely to be a valid assumption in randomised trials, and in this simulation study independence was ensured through stratified treatment allocation. Further simulation studies are needed to assess the impact of substantial imbalance in treatment allocation across covariate categories.

We estimated $\tau^2$ via restricted maximum likelihood as recommended for one- and two-stage models when random treatment effects are assumed.[29, 30] DerSimonian and Laird's method was commonly used for two-stage models in the past.[31] Levels of agreement between values of $I^2$ derived from the one- and two-stage models would likely not have been as high if this method had been used for the two-stage model due to potential downward bias of $\tau^2$.[29] Methods for calculating $I^2$ for aggregate data fixed-effect MAs require evaluation of Cochrane's $Q$ statistic.[4] Cochrane's $Q$ is derived using trial-specific treatment effects which are not directly estimated in one-stage IPD-MA.[32] Thus, additional work is required to produce an equivalent to $I^2$ from a one-stage fixed-effect IPD-MA.

Cochrane's $Q$ statistic is also often utilised when deriving 95% CI around $I^2$ (uncertainty intervals),[4, 17] which we have not investigated in this study. 95% CI around $\hat{\tau}^2$ are readily obtained following IPD-MA, although values can differ substantially depending on which method has been used to derive the interval. Uncertainty around the average within-trial variance ($\bar{\sigma}_\theta^2$) also needs to be considered when reporting intervals for $I^2$ in one-stage IPD-MA (particularly when sample sizes are small).[17] Bootstrapping procedures may be one way to approach this in the future.[33]

In this study, models were assessed under the frequentist framework and additional considerations may be needed for Bayesian MA.

### 5.3   Conclusions

$I^2$ can be derived for one-stage random effects IPD-MA of treatment-covariate interactions with continuous outcomes to quantify between-trial heterogeneity. Careful consideration should be given to the validity of modelling assumptions prior to interpreting $I^2$ regardless of whether a one- or two-stage model is undertaken.

**References**

1.	Wang H, Chen Y, Lin Y, et al. The methodological quality of individual participant data meta-analysis on intervention effects: systematic review. *BMJ* 2021; 373: n736. 20210419. DOI: 10.1136/bmj.n736.
2.	Higgins JPT. Commentary: Heterogeneity in meta-analysis should be expected and appropriately quantified. *Int J Epidemiol* 2008; 37: 1158-1160. DOI: 10.1093/ije/dyn204.
3.	Stewart LA, Clarke M, Rovers M, et al. Preferred Reporting Items for Systematic Review and Meta-Analyses of individual participant data: the PRISMA-IPD Statement. *JAMA* 2015; 313: 1657-1665. 2015/04/29. DOI: 10.1001/jama.2015.3656.
4.	Higgins JPT and Thompson SG. Quantifying heterogeneity in a meta-analysis. *Stat Med* 2002; 21: 1539-1558. DOI: 10.1002/sim.1186.
5.	Hemming K, Hughes JP, McKenzie JE, et al. Extending the I-squared statistic to describe treatment effect heterogeneity in cluster, multi-centre randomized trials and individual patient data meta-analysis. *Stat Methods Med Res* 2021; 30: 376-395. 20200921. DOI: 10.1177/0962280220948550.
6.	Riley RD, Debray TPA, Fisher D, et al. Individual participant data meta-analysis to examine interactions between treatment effect and participant-level covariates: Statistical recommendations for conduct and planning. *Stat Med* 2020; 39: 2115-2137. 20200430. DOI: 10.1002/sim.8516.
7.	Simmonds MC and Higgins JPT. Covariate heterogeneity in meta-analysis: Criteria for deciding between meta-regression and individual patient data. *Stat Med* 2007; 26: 2982-2999. DOI: 10.1002/sim.2768.
8.	Hattle M, Ensor J, Scandrett K, et al. Individual participant data meta-analysis to examine linear or non-linear treatment-covariate interactions at multiple time-points for a continuous outcome. *Res Synth Methods* 2024; Epub 20240916. DOI: 10.1002/jrsm.1750.
9.	Wu Z, Crabb DP, Chauhan BC, et al. Improving the feasibility of glaucoma clinical trials using trend-based visual field progression end points. *Ophthalmol Glaucoma* 2019; 2: 72-77. 20190117. DOI: 10.1016/j.ogla.2019.01.004.
10.	Levin LA, Sengupta M, Balcer LJ, et al. Report from the National Eye Institute workshop on neuro-ophthalmic disease clinical trial endpoints: optic neuropathies. *Invest Ophthalmol Vis Sci* 2021; 62: 30. 2021/12/01. DOI: 10.1167/iovs.62.14.30.
11.	McGuinness MB, Yelland LN and Hui F. Sample size reassessment for a randomised clinical trial with visual field outcomes correlated between eyes, visits, and repeated tests within visits. *ACTA National Conference*. Adelaide, SA 2022.
12.	Otarola F, Chen A, Morales E, et al. Course of glaucomatous visual field loss across the entire perimetric range. *JAMA Ophthalmology* 2016; 134: 496-502. DOI: 10.1001/jamaophthalmol.2016.0118.
13.	Burke DL, Ensor J and Riley RD. Meta-analysis using individual participant data: one-stage and two-stage approaches, and why they may differ. *Stat Med* 2017; 36: 855-875. 20161016. DOI: 10.1002/sim.7141.
14.	McGuinness MB, McKenzie JE, Hui F, et al. Quantifying inconsistency in one-stage individual participant data meta-analyses of treatment-covariate interactions: a simulation study protocol. *OSF Registries* 2024 JUNE 26. DOI: 10.17605/OSF.IO/W3SQE.
15.	Matsumoto M and Nishimura T. Mersenne twister: a 623-dimensionally equidistributed uniform pseudo-random number generator. *ACM Trans Model Comput Simul* 1998; 8: 3–30. DOI: 10.1145/272991.272995.
16.	Fisher DJ. Two-stage individual participant data meta-analysis and generalized forest plots. *Stata J* 2015; 15: 369-396. DOI: 10.1177/1536867x1501500203.
17.	Viechtbauer W. Confidence intervals for the amount of heterogeneity in meta-analysis. *Stat Med* 2007; 26: 37-52. DOI: 10.1002/sim.2514.

18.	Thorlund K, Imberger G, Johnston BC, et al. Evolution of heterogeneity (I2) estimates and their 95% confidence intervals in large meta-analyses. *PLoS ONE* 2012; 7: e39471. DOI: 10.1371/journal.pone.0039471.
19.	Mittlböck M and Heinzl H. A simulation study comparing properties of heterogeneity measures in meta-analyses. *Stat Med* 2006; 25: 4321-4333. DOI: 10.1002/sim.2692.
20.	Morris TP, Fisher DJ, Kenward MG, et al. Meta-analysis of Gaussian individual patient data: Two-stage or not two-stage? *Stat Med* 2018; 37: 1419-1438. 20180118. DOI: 10.1002/sim.7589.
21.	Riley RD, Ensor J, Hattle M, et al. Two-stage or not two-stage? That is the question for IPD meta-analysis projects. *Res Synth Methods* 2023; 14: 903-910. 20230822. DOI: 10.1002/jrsm.1661.
22.	Stijnen T, Hamza TH and Ozdemir P. Random effects meta-analysis of event outcome in the framework of the generalized linear mixed model with applications in sparse data. *Stat Med* 2010; 29: 3046-3067. DOI: 10.1002/sim.4040.
23.	Ades AE, Lu G and Higgins JPT. The interpretation of random-effects meta-analysis in decision models. *Med Decis Making* 2005; 25: 646-654. DOI: 10.1177/0272989X05282643.
24.	Rucker G, Schwarzer G, Carpenter JR, et al. Undue reliance on I2 in assessing heterogeneity may mislead. *BMC Med Res Methodol* 2008; 8: 79. 20081127. DOI: 10.1186/1471-2288-8-79.
25.	Ioannidis JPA, Patsopoulos NA and Evangelou E. Uncertainty in heterogeneity estimates in meta-analyses. *BMJ* 2007; 335: 914-916. DOI: 10.1136/bmj.39343.408449.80.
26.	Jackson D, White IR and Riley RD. Quantifying the impact of between-study heterogeneity in multivariate meta-analyses. *Stat Med* 2012; 31: 3805-3820. 20120704. DOI: 10.1002/sim.5453.
27.	Kontopantelis E and Reeves D. A short guide and a forest plot command (ipdforest) for one-stage meta-analysis. *Stata J* 2013; 13: 574-587. DOI: 10.1177/1536867x1301300308.
28.	Riley RD, Ensor J, Jackson D, et al. Deriving percentage study weights in multi-parameter meta-analysis models: with application to meta-regression, network meta-analysis and one-stage individual participant data models. *Stat Methods Med Res* 2018; 27: 2885-2905. 20170206. DOI: 10.1177/0962280216688033.
29.	Langan D, Higgins JPT, Jackson D, et al. A comparison of heterogeneity variance estimators in simulated random-effects meta-analyses. *Res Synth Methods* 2019; 10: 83-98. 20180906. DOI: 10.1002/jrsm.1316.
30.	Riley RD, Legha A, Jackson D, et al. One-stage individual participant data meta-analysis models for continuous and binary outcomes: Comparison of treatment coding options and estimation methods. *Stat Med* 2020; 39: 2536-2555. 20200511. DOI: 10.1002/sim.8555.
31.	DerSimonian R and Laird N. Meta-analysis in clinical trials. *Control Clin Trials* 1986; 7: 177-188. DOI: 10.1016/0197-2456(86)90046-2.
32.	Cochran WG. The combination of estimates from different experiments. *Biometrics* 1954; 10: 101-129. DOI: 10.2307/3001666.
33.	Turner RM, Omar RZ, Yang M, et al. A multilevel model framework for meta-analysis of clinical trials with binary outcomes. *Stat Med* 2000; 19: 3417-3432. DOI: 10.1002/1097-0258(20001230)19:24<3417::aid-sim614>3.0.co;2-l.

**Supplemental material**

Supplement A: Derivation of within-trial treatment-covariate interaction effect variance ($\sigma^2_{\theta j}$) from a one-stage model.

Supplement B: Example Stata code to generate and analyse data.

Supplement C: Supplementary tables and figures.

**Supplemental Material A: Derivation of within-trial treatment-covariate interaction effect variance ($\sigma^2_{\theta j}$) from a one-stage model.**

In a random-effects meta-analysis aiming to investigate the difference in treatment effect between two categories of a binary subgroup variable using data collected at the final study visit (i.e., one observation per participant), the estimate of interest is the coefficient for the treatment-subgroup interaction term. A model for $y_{ij}$, the outcome variable (visual field mean deviation) for individual $i$ from trial $j$, could be:

$$y_{ij} = (\beta_0 + u_{0j}) + a_{ij}z_{ij}(\beta_1 + u_{1j}) + \beta_2 a_{ij} + \beta_3 z_{ij} + e_{ij} \tag{A1}$$

where:

$\beta_1$ represents the true treatment-subgroup interaction effect across all trials ($\theta$),

$a_{ij}$ is the binary treatment group indicator (0 = control, 1 = active treatment),

$z_{ij}$ is the binary glaucoma subgroup indicator (0 = primary open angle, 1 = pseudoexfoliative glaucoma),

$u_{0j}$ is the trial-specific fixed intercept,

$u_{1j}$ is the random trial-specific treatment-subgroup interaction effect with distribution $\sim^{\text{iid}} N[0, \tau^2_{u1}]$,

and $e_{ij}$ is the random model error/residual with distribution $\sim^{\text{iid}} N(0, \sigma^2_{ej})$.

If $\bar{Y}_{.jZ=z,A=a}$ represents the average outcome for the $n_{jza}$ participants with subgroup $z$ in treatment group $a$ of trial $j$, the treatment-subgroup interaction effect for trial $j$ is:

$$\theta_j = (\bar{Y}_{.jZ=1,A=1} - \bar{Y}_{.jZ=1,A=0}) - (\bar{Y}_{.jZ=0,A=1} - \bar{Y}_{.jZ=0,A=0}) \tag{A2}$$

where:

$\bar{Y}_{.jZ=1,A=1} = (\beta_0 + u_{0j}) + (\beta_1 + u_{1j}) + \beta_2 + \beta_3 + \bar{e}_{.jZ=1,A=1}$,

$\bar{Y}_{.jZ=1,A=0} = (\beta_0 + u_{0j}) + \beta_3 + \bar{e}_{.jZ=1,A=0}$,

$\bar{Y}_{.jZ=0,A=1} = (\beta_0 + u_{0j}) + \beta_2 + \bar{e}_{.jZ=0,A=1}$, and

$\bar{Y}_{.jZ=0,A=0} = (\beta_0 + u_{0j}) + \bar{e}_{.jZ=0,A=0}$.

Thus:

$$\theta_j = \beta_1 + u_{1j} + \bar{e}_{.jZ=1,A=1} - \bar{e}_{.jZ=1,A=0} - \bar{e}_{.jZ=0,A=1} + \bar{e}_{.jZ=0,A=0} \tag{A3}$$

Assuming that the variance of the error terms ($\sigma^2_{ej}$) is the same in each subgroup and treatment arm, the variance of the treatment-covariate interaction effect for trial $j$ is:

$$\sigma^2_{\theta j} = var(\bar{e}_{.jZ=1,A=1} - \bar{e}_{.jZ=1,A=0} - \bar{e}_{.jZ=0,A=1} + \bar{e}_{.jZ=0,A=0}) \tag{A4}$$

$$= var(\bar{e}_{.jZ=1,A=1}) + var(\bar{e}_{.jZ=1,A=0}) + var(\bar{e}_{.jZ=0,A=1}) + var(\bar{e}_{.jZ=0,A=0})$$

$$= var\left(\frac{\sum_{i=1}^{n_j} e_{ijZ=1,A=1}}{n_{j11}}\right) + var\left(\frac{\sum_{i=1}^{n_j} e_{ijZ=1,A=0}}{n_{j10}}\right) + var\left(\frac{\sum_{i=1}^{n_j} e_{ijZ=0,A=1}}{n_{j01}}\right) + var\left(\frac{\sum_{i=1}^{n_j} e_{ijZ=0,A=0}}{n_{j00}}\right)$$

$$= \frac{\sigma^2_{ej}}{n_{j11}} + \frac{\sigma^2_{ej}}{n_{j10}} + \frac{\sigma^2_{ej}}{n_{j01}} + \frac{\sigma^2_{ej}}{n_{j00}}$$

Given disease subgroup is a binary variable, the proportion of participants in trial $j$ with subgroup $Z = z$, is equal to the number in that subgroup divided by the total number in the trial $\left(p_{jZ=z} = \frac{n_{jZ=z}}{n_j}\right)$, and the sample variance of $Z$, $var(Z_j)$, is equal to $p_{jZ=0}p_{jZ=1}$. Likewise, the proportion of people with treatment group $A = a$ is $p_{jA=a}$, and $var(A_j) = p_{jA=0}p_{jA=1}$.

The number of participants in trial $j$ with subgroup $Z = z$ in treatment group $A = a$ is equal to the total number of participants in the trial multiplied by the probability that both $Z = z$ and $A = a$, i.e., $n_{jza} = n_j \times P(Z = z \cap A = a)$. When treatment group allocation is stratified by disease subgroup so that the probability that $A = 1$ is equal in each subgroup (i.e., $A$ is independent of $Z$), then $P(Z = z \cap A = a) = P(Z = z)P(A = a)$. Thus, $n_{jza}$ is equal to $n_j \times p_{jZ=z} \times p_{jA=a}$.

Therefore, Equation A4 simplifies to:

$$\sigma^2_{\theta j} = \frac{\sigma^2_{ej}}{n_j}\left(\frac{1}{p_{jZ=1}p_{jA=1}} + \frac{1}{p_{jZ=1}p_{jA=0}} + \frac{1}{p_{jZ=0}p_{jA=1}} + \frac{1}{p_{jZ=0}p_{jA=0}}\right) \tag{A5}$$

$$= \frac{\sigma^2_{ej}}{n_j}\left(\frac{1}{p_{jZ=0}p_{jZ=1}} + \frac{1}{p_{jA=0}p_{jA=1}}\right)$$

$$= \frac{\sigma^2_{ej}}{n_j var(Z_j) var(A_j)}$$

**Supplemental Material B: Example Stata code to generate and analyse data**

```stata
version 18

*-----------------------------------------------------------------------
**# Install user-written commands if needed
*-----------------------------------------------------------------------

foreach command in ereplace metan ipdmetan {
      capture noisily which `command' // gives current version installed
      if _rc == 111 { // capture return code = 111 if not installed
            ssc install `command'
      }
}

*-----------------------------------------------------------------------
**# Program to generate single dataset
*-----------------------------------------------------------------------

/* Each participant has one observation per visit and multiple visits

* Syntax options
num_trials   number of trials in meta-analyses
sigma2_u1    variance of the random trial-specific treatment-time-subgroup
             interaction effect (tau2 for Estimand 2)
np           average number of people in one treatment arm of each trial
pr_z         proportion of people with glaucoma subgroup variable z=1
sdpint       standard deviation of the random participant-level intercept
             error
num_visits   number of visits per participant
sde          standard deviation of the residual model error
beta1        treatment-covariate-time interaction effect (theta for
             Estimand2)

* Generated variables
k            trial number
trialj       binary indicator for trial j
id           participant ID number
z            binary glaucoma subgroup
a            binary treatment group
t            time from baseline in years
y            outcome variable, visual field mean deviation in decibels
y0           value of outcome variable at baseline
tx_time      treatment-time interaction
tx_z         treatment-subgroup interaction
z_time       subgroup-time interaction
tx_z_time    treatment-subgroup-time interaction
var_trij     trial-specific variables: equal to var if trial = j, 0
             otherwise */
```

```stata
capture program drop sim_ipd_ma_i2_generate
program define sim_ipd_ma_i2_generate

    version 18

    syntax, [num_trials(real 4) sigma2_u1(real 0.01) ///
      np(real 200) sde(real 1) pr_z(real 0.375) ///
      SDPInt(real 2) num_visits(real 5) beta1(real 0.12)]

    tempvar ss u_k_slope u_p_int e

    drop _all
    qui set obs `num_trials'

    gen k = _n // trial ID
    qui tab k, gen(trial) // Create dummy variable for each trial

    local sigma_u1 = sqrt(`sigma2_u1')
    // sigma2_u1 represents variance of random tx-subgroup-time effect for
    Estimand 2, need SD (not variance) for rnormal command
    gen `u_k_slope' = rnormal(0,`sigma_u1') // Slope error for each trial

    gen `ss' = rnormal(`np',10)
    // Number of people in in each tx arm, different for each trial
    qui expand `ss' // Create 1 obs per participant in treatment arm

    gen z = uniform() < `pr_z'
    // Randomly allocate binary glaucoma subgroup with given probability
    qui expand 2 // total samp size/trial = 2 x n in each arm
    bysort k z: gen a = (_n > _N/2)
    // Half in each glaucoma subgroup within each trial have active
    treatment

    gen id = _n // unique person ID
    gen `u_p_int' = rnormal(0,`sdpint') // Person intercept error

    qui expand `num_visits' // Create one observation per visit
    bysort id: gen t = (_n - 1) * 0.5
    // Generate time variable, review every 6 months (0.5 years)
    gen `e' = rnormal(0,`sde') // Residual model error

    *   Generate outcome variable (visual field MD) at each visit
    gen y = round((`beta1' + `u_k_slope')*(t * a * z) + (-0.4 * t) ///
        + (0.12 * a * t) + `u_p_int' + `e', 0.01)
    qui gen y0 = y if t == 0 // baseline VF MD
    qui bysort id: ereplace y0 = min(y0)

    *   Generate treatment-covariate interaction variables
    gen tx_time = a * t
    gen tx_z = a * z
    gen z_time = z * t
    gen tx_z_time = a * z * t

    *   Generate separate baseline, subgroup, tx & interaction variables for
    each trial
    foreach j of numlist 1/`num_trials' {
        gen tx_tri`j'        = trial`j' * a
        gen z_tri`j'         = trial`j' * z
        gen y0_tri`j'        = trial`j' * y0
        gen time_tri`j'      = trial`j' * t
        gen tx_time_tri`j'   = trial`j' * a * t
```

```stata
            gen tx_z_tri`j'         = trial`j' * a * z
            gen z_time_tri`j'       = trial`j' * z * t
        }

end

*-----------------------------------------------------------------------------
**# Analyse single IPD dataset
*-----------------------------------------------------------------------------

/* Each estimand is assessed by three models.

Estimand 1: Treatment-subgroup interaction at final study visit.
Estimand 2: Treatment-subgroup-time interaction across repeated visits.

M2:     Two-stage model.
M1s:    Full one-stage model with residual error variance estimated
        separately for each trial.
M1c:    Parsimonious one-stage model with common residual error variance  */

** Write program to calculate average within-trial variance & estimate I2
*-----------------------------------------------------------------------------

/* Matrix specified in the "`weight_mat'" option must have:
    -   the number of rows equal to the number of trials,
    -   the fixed-effect inverse variance weights for each trial in the
        first column,
    -   the square of those weights in the second column.
Trial weights derived separately for each model below.              */

capture program drop sim_ipd_ma_i2_derive
program define sim_ipd_ma_i2_derive

    version 18

    syntax, weight_mat(string)

    tempname sum_col wf_sum wf2_sum awv

    mata: st_matrix("`sum_col'", colsum(st_matrix("`weight_mat'")))
    // creates row vector called sum_col with the sum of values in each
    column
    scalar `wf_sum'  = `sum_col'[1,1] // sum of the trial weights
    scalar `wf2_sum' = `sum_col'[1,2] // sum of the squared trial weights

    scalar `awv' = (`=total_trials' - 1) / ///
        (`wf_sum' - (`wf2_sum' / `wf_sum')) // Equation 3

    disp _newline "{hline 60}"
    disp "Sum of weights = " round(`wf_sum',0.01) ///
        ", sum of squared weights = " round(`wf2_sum',0.01)
    disp "Average within-trial interaction effect variance = " ///
        round(`awv',.0001)
    disp "Estimated tau2 = " round(tau2_hat, 0.001)
    disp "Estimated I-squared = " ///
        round(100 * tau2_hat / (tau2_hat + `awv'),0.1) "%" // Equation 1
    disp "{hline 60}"

end
```

```
**      Generate a single dataset using program specified above
*------------------------------------------------------------
set seed 11111 // specify starting seed
sim_ipd_ma_i2_generate,
// Change scenario factors using options specified above

**      Create scalar to record the number of trials in the meta-analysis
*------------------------------------------------------------------------
qui sum k // k = trial number
scalar total_trials = r(max)

**      Number of obs & variance of variables in contrast for each trial
*----------------------------------------------------------------------
/* Assumes non-missing values for all variables in contrast
Estimate pop variance as (sum of squares)/N
        rather than using sample variance as returned by the summarize
        command which uses a divisor of (N-1)                      */

*       Estimand 1
foreach j of numlist 1 /`=total_trials' {
    disp _newline "Estimand 1, Trial `j':"

    foreach v of varlist a z { // Treatment group & subgroup
        qui regress `v' if k == `j' & t == 0 // distribution at baseline
        scalar var_`v'`j'_e1 = e(rss) / e(N)
        disp "      Variance(`v') = " var_`v'`j'_e1
    }

    scalar num_obs`j'_e1 = e(N)
    disp "   Number of observations = " num_obs`j'_e1
}

*       Estimand 2
foreach j of numlist 1 /`=total_trials' {
    disp _newline "Estimand 2, Trial `j':"

    foreach v of varlist a z t { // Treatment group, subgroup, & time
        qui regress `v' if k == `j' // distribution across all visits
        scalar var_`v'`j'_e2 = e(rss) / e(N)
        disp "      Variance(`v') = " var_`v'`j'_e2

    }

    scalar num_obs`j'_e2 = e(N)
    disp "   Number of observations = " num_obs`j'_e2
}
```

```
**      Conduct meta-analyses for Estimand 1
*----------------------------------------

*       Estimand 1, 2-stage model (M2)
ipdmetan, study(k) keepall nograph re(reml): ///
   regress y i.(tx_z i.a i.z) c.y0 if t == 2
scalar tau2_hat = r(tau2)
matrix A = r(coeffs) // matrix with trial-level SE of tx-cov effect
disp "Average within-trial interaction effect variance from ipdmetan
command: " r(sigmasq)

matrix W = J(total_trials,2,.)
// Create an empty matrix with one row per trial to store trial weights
foreach j of numlist 1/ `=total_trials' {
   scalar w = 1 / A["r`j'","_seES"]^2
   // weight = 1/variance[interaction effect]
   matrix W[`j',1] = w // Fixed-effect inverse variance weight in 1st
   column
   matrix W[`j',2] = w^2 // Squared weight in 2nd column
}
matrix list W

sim_ipd_ma_i2_derive, weight_mat(W)
/* Call program defined above to estimate average within-trial variance
        and I2 using inverse variance weights stored in matrix W */

*       Estimand 1, full 1-stage model (M1s)
/* Use residual model error variance ("j: var(e)") in calculation to
approximate variance of the interaction effect separately for each trial */

mixed y tx_z i.(tx_tri* z_tri* trial*) c.(y0_tri*) if t == 2 ///
      || k: tx_z, nocons reml iterate(40) nolog residuals(ind, by(k))
// "by(k)" gives residual error variance separately for each trial
scalar tau2_hat = r(table)["b","k:var(tx_z)"]
matrix E = r(table) // Matrix includes residual error variance estimates

matrix W = J(total_trials,2,.) // Empty matrix for weights
foreach j of numlist 1/`=total_trials' {
   scalar w = (num_obs`j'_e1 * var_a`j'_e1 * var_z`j'_e1) / ///
      (E["b","Residual:var(e)#`j'.k"])
   // weight = 1/variance[interaction effect]

   matrix W[`j',1] = w // Fixed-effect inverse variance weight
   matrix W[`j',2] = w^2 // Squared weight
}
matrix list W

sim_ipd_ma_i2_derive, weight_mat(W) // Estimate I2 using weights stored in
W
```

```
*       Estimand 1, parsimonious 1-stage model (M1c)
/* Use common residual model error variance ("var(Residual)") in
calculation to approximate variance of the interaction effect for all
trials */

mixed y tx_z i.(tx_tri* z_tri* trial*) c.(y0_tri*) if t == 2 ///
      || k: tx_z, nocons reml iterate(40) nolog residuals(ind)
scalar tau2_hat = r(table)["b","k:var(tx_z)"]
matrix E = r(table) // Matrix includes residual error variance estimate

matrix W = J(total_trials,2,.) // Empty matrix for weights
foreach j of numlist 1/`=total_trials' {
   scalar w = (num_obs`j'_e1 * var_a`j'_e1 * var_z`j'_e1) / ///
      E["b","Residual:var(e)"]  // weight = 1/variance[interaction effect]

   matrix W[`j',1] = w // Fixed-effect inverse variance weight
   matrix W[`j',2] = w^2 // Squared weight
}
matrix list W

sim_ipd_ma_i2_derive, weight_mat(W) // Estimate I2 using weights stored in
W

**     Conduct meta-analyses for Estimand 2
*----------------------------------------

* Estimand 2, 2-stage model (M2)
ipdmetan, study(k) keepall nograph re(reml): ///
      mixed y c.a#c.z#c.t c.tx_time c.z_time c.t i.a i.z i.tx_z || id:,
reml
scalar tau2_hat = r(tau2)
matrix A = r(coeffs) // matrix with trial-level SE of tx-cov effect
disp "Average within-trial interaction effect variance from ipdmetan
command: " r(sigmasq)

matrix W = J(total_trials,2,.) // Empty matrix for weights
foreach j of numlist 1/`=total_trials' {
   scalar w = 1 / A["r`j'","_seES"]^2
   // weight = 1/variance[interaction effect]

   matrix W[`j',1] = w // Fixed-effect inverse variance weight
   matrix W[`j',2] = w^2 // Squared weight
}
matrix list W

sim_ipd_ma_i2_derive, weight_mat(W) //Estimate I2 using weights stored in W
```

```
* Estimand 2, full 1-stage model (M1s)
/* Use residual model error variance ("j: var(e)") in calculation to
approximate variance of the interaction effect separately for each trial */

mixed y c.(tx_z_time tx_time_tri* z_time_tri* time_tri*) ///
   i.(tx_tri* z_tri* tx_z_tri* trial*) ///
   || k: tx_z_time, nocons || id:, ///
   reml iterate(40) nolog residuals(ind, by(k))
scalar tau2_hat = r(table)["b","k:var(tx_z_time)"]
matrix E = r(table) // Matrix includes residual error variance estimates

matrix W = J(total_trials,2,.) // Empty matrix for weights
foreach j of numlist 1/`=total_trials' {
   scalar w = (num_obs`j'_e2 * var_a`j'_e2 * var_z`j'_e2 * var_t`j'_e2) /
   ///
      E["b","Residual:var(e)#`j'.k"]
   // weight = 1/variance[interaction effect]

   matrix W[`j',1] = w // Fixed-effect inverse variance weight
   matrix W[`j',2] = w^2 // Squared weight
}
matrix list W

sim_ipd_ma_i2_derive, weight_mat(W) //Estimate I2 using weights stored in W

* Estimand 2, parsimonious 1-stage model (M1c)
/* Use common residual model error variance ("var(Residual)") in
calculation to approximate variance of the interaction effect for all
trials */

mixed y c.(tx_z_time tx_time_tri* z_time_tri* time_tri*) ///
   i.(tx_tri* z_tri* tx_z_tri* trial*) ///
   || k: tx_z_time, nocons || id:, ///
   reml iterate(40) nolog residuals(ind)
scalar tau2_hat = r(table)["b","k:var(tx_z_time)"]
matrix E = r(table) // Matrix includes residual error variance estimate

matrix W = J(total_trials,2,.) // Empty matrix for weights
foreach j of numlist 1/`=total_trials' {
   scalar w = (num_obs`j'_e2 * var_a`j'_e2 * var_z`j'_e2 * var_t`j'_e2) /
   ///
      E["b","Residual:var(e)"]  // weight = 1/variance[interaction effect]

   matrix W[`j',1] = w // Fixed-effect inverse variance weight
   matrix W[`j',2] = w^2 // Squared weight
}
matrix list W

sim_ipd_ma_i2_derive, weight_mat(W) //Estimate I2 using weights stored in W
```

**Supplemental Material C: Supplementary tables and figures**

Supplemental Table 1: Percentage of one-stage models that converged within 40 iterations.

|          |        |             |                | % of models converged | | | |
|----------|--------|-------------|----------------|-------------|------|-------------|------|
|          |        |             |                | **Estimand 1** | | **Estimand 2** | |
| **Scenario** | **Trials** | $\bar{n}_j$ | **Heterogeneity** | M1c | M1s | M1c | M1s |
| 1  | 4  | 90  | None | 99.8  | 99.7  | 99.2 | 99.5 |
| 2  |    |     | Low  | 99.9  | 99.8  | 99.1 | 98.9 |
| 3  |    |     | High | 99.9  | 99.9  | 99.2 | 99.0 |
| 4  |    | 400 | None | 99.9  | 98.7  | 97.4 | 96.9 |
| 5  |    |     | Low  | 99.6  | 99.2  | 97.9 | 97.7 |
| 6  |    |     | High | 99.6  | 99.9  | 98.5 | 98.4 |
| 7  | 16 | 90  | None | 99.2  | 99.6  | 97.8 | 98.5 |
| 8  |    |     | Low  | 99.3  | 99.2  | 98.3 | 96.7 |
| 9  |    |     | High | 99.7  | 99.9  | 99.1 | 99.0 |
| 10 |    | 400 | None | 97.0  | 97.4  | 93.8 | 94.8 |
| 11 |    |     | Low  | 99.3  | 99.3  | 96.4 | 97.3 |
| 12 |    |     | High | 99.9  | 99.9  | 99.9 | 99.9 |
| 13 | 30 | 90  | High | 99.8  | 99.6  | NA   | NA   |
| 14 |    | 400 | High | 100.0 | 100.0 | NA   | NA   |

1000 datasets in each scenario; likelihood maximisation capped at 40 iterations; all models produced estimates in all datasets. Estimand 1 = treatment-covariate interaction at two years, Estimand 2 = treatment-covariate-time interaction. M1c = parsimonious 1-stage model with common residual error variance. M1s = full 1-stage model with trial-level residual error variance, $\bar{n}_j$ = average number of participants per trial. Only Estimand 1 assessed in scenarios with 30 trials.

Supplemental Table 2: Median time to run models.

| | | | | Median [IQR] time (seconds) | | |
|---|---|---|---|---|---|---|
| **Scenario** | **Trials** | $\bar{n}_j$ | **Heterogeneity** | **Parsimonious (M1c)** | **Full (M1s)** | **Two-stage (M2)** |
| **Estimand 1** | | | | | | |
| 1 | 4 | 90 | None | 0.7 [ 0.6, 0.8] | 1.5 [ 1.4, 1.7] | 0.2 [ 0.2, 0.3] |
| 2 | | | Low | 0.6 [ 0.5, 0.8] | 1.5 [ 1.4, 1.7] | 0.2 [ 0.2, 0.3] |
| 3 | | | High | 0.6 [ 0.5, 0.7] | 1.6 [ 1.4, 1.7] | 0.2 [ 0.2, 0.3] |
| 4 | | 400 | None | 1.2 [ 0.9, 1.4] | 2.7 [ 2.4, 3.1] | 0.2 [ 0.2, 0.3] |
| 5 | | | Low | 1.0 [ 0.8, 1.3] | 2.7 [ 2.4, 3.0] | 0.2 [ 0.2, 0.3] |
| 6 | | | High | 0.8 [ 0.7, 1.0] | 2.4 [ 2.2, 2.7] | 0.2 [ 0.2, 0.3] |
| 7 | 16 | 90 | None | 2.2 [ 1.7, 2.7] | 34.5 [31.4,37.8] | 0.3 [ 0.3, 0.4] |
| 8 | | | Low | 2.0 [ 1.7, 2.7] | 34.6 [31.4,38.6] | 0.3 [ 0.3, 0.4] |
| 9 | | | High | 1.7 [ 1.5, 1.9] | 33.5 [30.9,36.4] | 0.3 [ 0.3, 0.4] |
| 10 | | 400 | None | 3.3 [ 2.5, 4.4] | 54.1 [45.6,69.3] | 0.4 [ 0.3, 0.5] |
| 11 | | | Low | 2.6 [ 2.1, 3.3] | 54.9 [47.9,69.8] | 0.4 [ 0.3, 0.5] |
| 12 | | | High | 2.4 [ 2.1, 2.9] | 62.5 [53.3,71.8] | 0.5 [ 0.4, 0.6] |
| 13 | 30 | 90 | High | 4.3 [ 4.0, 4.6] | 277.0 [260.8,304.3] | 0.6 [ 0.6, 0.7] |
| 14 | | 400 | High | 8.3 [ 7.0,10.3] | 851.4 [691.1,977.2] | 1.0 [ 0.9, 1.2] |
| **Estimand 2** | | | | | | |
| 1 | 4 | 90 | None | 2.9 [ 2.1, 3.6] | 6.1 [ 5.4, 6.8] | 1.9 [ 1.8, 2.1] |
| 2 | | | Low | 2.8 [ 2.1, 3.5] | 6.0 [ 5.4, 6.7] | 1.9 [ 1.7, 2.1] |
| 3 | | | High | 2.3 [ 2.0, 3.3] | 5.7 [ 5.2, 6.5] | 2.0 [ 1.8, 2.1] |
| 4 | | 400 | None | 6.7 [ 5.3, 8.4] | 13.5 [11.5,15.4] | 3.5 [ 3.3, 3.8] |
| 5 | | | Low | 5.9 [ 5.0, 8.2] | 12.5 [10.9,15.1] | 3.8 [ 3.6, 4.0] |
| 6 | | | High | 4.9 [ 4.4, 5.5] | 11.0 [10.0,12.8] | 3.6 [ 3.4, 3.8] |
| 7 | 16 | 90 | None | 15.3 [11.9,20.6] | 197.8 [184.1,215.8] | 6.0 [ 5.6, 6.3] |
| 8 | | | Low | 13.2 [11.5,19.5] | 196.4 [182.3,215.0] | 6.2 [ 5.8, 6.5] |
| 9 | | | High | 11.7 [10.9,12.9] | 195.1 [183.7,210.0] | 6.1 [ 5.7, 6.5] |
| 10 | | 400 | None | 61.7 [43.8,79.8] | 723.9 [606.3,928.4] | 10.8 [10.0,13.1] |
| 11 | | | Low | 44.7 [36.0,54.9] | 721.7 [602.4,849.6] | 11.0 [10.1,13.0] |
| 12 | | | High | 41.6 [35.8,47.8] | 744.4 [613.4,888.8] | 12.8 [11.5,14.6] |

Likelihood maximisation capped at 40 iterations. Estimand 1 = treatment-covariate interaction at two years, Estimand 2 = treatment-covariate-time interaction. M1c = parsimonious 1-stage model with common residual error variance, M1s = full 1-stage model with trial-level residual error variance, M2 = 2-stage model, $\bar{n}_j$ = average number of participants per trial. Only Estimand 1 assessed in scenarios with 30 trials.

Supplemental Table 3: Bias, empirical standard error, and mean square error for $I^2$ (percentage points).

| Het | $\bar{n}_j$ | Trials | True $I^2$ (%) | Parsimonious (M1c) | | | | Full (M1s) | | | | Two-stage (M2) | | | |
|---|---|---|---|---|---|---|---|---|---|---|---|---|---|---|---|
| | | | | Mean | Bias (MCSE) | ESE | MSE | Mean | Bias (MCSE) | ESE | MSE | Mean | Bias (MCSE) | ESE | MSE |
| **Estimand 1** | | | | | | | | | | | | | | | |
| None | 90 | 4 | 0.0 | 14.3 | 14.3 (0.7) | 22.6 | 714.1 | 14.7 | 14.7 (0.7) | 22.8 | 737.1 | 14.9 | 14.9 (0.7) | 22.8 | 739.7 |
| | | 16 | | 10.0 | 10.0 (0.5) | 14.4 | 306.3 | 10.4 | 10.4 (0.5) | 14.9 | 330.4 | 11.3 | 11.3 (0.5) | 15.2 | 357.7 |
| | 400 | 4 | 0.0 | 14.7 | 14.7 (0.7) | 21.8 | 691.0 | 14.7 | 14.7 (0.7) | 21.8 | 689.1 | 14.7 | 14.7 (0.7) | 21.8 | 689.4 |
| | | 16 | | 9.0 | 9.0 (0.4) | 14.0 | 276.3 | 9.0 | 9.0 (0.4) | 14.0 | 278.0 | 9.2 | 9.2 (0.4) | 14.1 | 282.4 |
| Low | 90 | 4 | 10.5 | 17.5 | 7.1 (0.7) | 23.5 | 600.2 | 17.6 | 7.1 (0.7) | 23.7 | 609.9 | 17.7 | 7.2 (0.7) | 23.6 | 609.4 |
| | | 16 | | 13.5 | 3.0 (0.5) | 16.2 | 271.4 | 14.0 | 3.5 (0.5) | 16.7 | 292.0 | 14.9 | 4.4 (0.5) | 16.9 | 303.4 |
| | 400 | 4 | 34.2 | 25.4 | -8.9 (0.9) | 28.5 | 891.3 | 25.5 | -8.8 (0.9) | 28.5 | 889.5 | 25.5 | -8.8 (0.9) | 28.5 | 887.7 |
| | | 16 | | 29.3 | -5.0 (0.6) | 20.5 | 443.6 | 29.3 | -5.0 (0.6) | 20.6 | 446.3 | 29.5 | -4.8 (0.6) | 20.5 | 440.8 |
| High | 90 | 4 | 36.9 | 27.7 | -9.3 (0.9) | 28.4 | 893.6 | 27.7 | -9.2 (0.9) | 28.5 | 896.0 | 27.8 | -9.1 (0.9) | 28.4 | 885.8 |
| | | 16 | | 32.3 | -4.7 (0.6) | 20.4 | 437.6 | 32.3 | -4.7 (0.7) | 21.0 | 463.0 | 33.2 | -3.7 (0.6) | 20.5 | 432.0 |
| | | 30 | | 32.5 | -4.4 (0.5) | 17.1 | 311.5 | 32.8 | -4.1 (0.6) | 17.6 | 327.5 | 33.8 | -3.1 (0.5) | 17.1 | 300.8 |
| | 400 | 4 | 72.3 | 52.0 | -20.3 (1.0) | 31.9 | 1427.2 | 52.0 | -20.2 (1.0) | 31.9 | 1423.9 | 52.0 | -20.2 (1.0) | 31.8 | 1422.3 |
| | | 16 | | 67.8 | -4.5 (0.4) | 13.6 | 206.2 | 67.8 | -4.4 (0.4) | 13.7 | 206.9 | 67.8 | -4.4 (0.4) | 13.6 | 205.5 |
| | | 30 | | 69.6 | -2.6 (0.3) | 8.6 | 80.7 | 69.7 | -2.5 (0.3) | 8.6 | 80.4 | 69.7 | -2.6 (0.3) | 8.6 | 80.0 |
| **Estimand 2** | | | | | | | | | | | | | | | |
| None | 90 | 4 | 0.0 | 14.0 | 14.0 (0.7) | 21.8 | 671.7 | 14.1 | 14.1 (0.7) | 21.9 | 680.1 | 14.2 | 14.2 (0.7) | 22.0 | 682.6 |
| | | 16 | | 9.9 | 9.9 (0.4) | 14.1 | 297.7 | 10.2 | 10.2 (0.5) | 14.2 | 306.1 | 10.4 | 10.4 (0.5) | 14.3 | 313.5 |
| | 400 | 4 | 0.0 | 13.3 | 13.3 (0.7) | 20.6 | 601.9 | 13.3 | 13.3 (0.7) | 20.6 | 601.3 | 13.3 | 13.3 (0.7) | 20.6 | 601.9 |
| | | 16 | | 9.3 | 9.3 (0.4) | 14.0 | 284.2 | 9.4 | 9.4 (0.4) | 14.1 | 286.0 | 9.5 | 9.5 (0.4) | 14.1 | 287.7 |
| Low | 90 | 4 | 11.6 | 17.0 | 5.3 (0.8) | 24.2 | 613.2 | 16.9 | 5.3 (0.8) | 24.3 | 618.7 | 17.0 | 5.4 (0.8) | 24.3 | 619.9 |
| | | 16 | | 14.2 | 2.6 (0.5) | 16.3 | 270.8 | 14.1 | 2.5 (0.5) | 16.4 | 274.6 | 14.4 | 2.8 (0.5) | 16.4 | 277.5 |
| | 400 | 4 | 36.9 | 26.9 | -10.1 (0.9) | 28.8 | 927.5 | 26.9 | -10.1 (0.9) | 28.7 | 925.6 | 26.9 | -10.0 (0.9) | 28.7 | 924.9 |
| | | 16 | | 31.3 | -5.7 (0.6) | 20.5 | 452.7 | 31.3 | -5.7 (0.7) | 20.6 | 454.1 | 31.3 | -5.6 (0.6) | 20.5 | 451.7 |
| High | 90 | 4 | 39.7 | 28.8 | -11.0 (0.9) | 28.3 | 921.2 | 28.7 | -11.0 (0.9) | 28.4 | 924.7 | 28.8 | -11.0 (0.9) | 28.4 | 922.9 |
| | | 16 | | 34.7 | -5.0 (0.6) | 20.3 | 435.2 | 34.7 | -5.1 (0.6) | 20.3 | 438.6 | 34.9 | -4.8 (0.6) | 20.2 | 431.9 |
| | 400 | 4 | 74.6 | 53.8 | -20.8 (1.0) | 31.7 | 1436.3 | 53.8 | -20.7 (1.0) | 31.7 | 1435.1 | 53.8 | -20.7 (1.0) | 31.7 | 1433.3 |
| | | 16 | | 70.3 | -4.3 (0.4) | 12.6 | 176.0 | 70.3 | -4.2 (0.4) | 12.6 | 176.9 | 70.3 | -4.2 (0.4) | 12.6 | 176.4 |

1000 simulated datasets per scenario. Estimand 1 = treatment-covariate interaction at two years, Estimand 2 = treatment-covariate-time interaction. ESE = empirical standard error, Het = between-trial heterogeneity, M1c = parsimonious 1-stage model with common residual error variance, M1s = full 1-stage model with trial-level residual error variance, M2 = 2-stage model, MCSE = Monte Carlo standard error, MSE = mean square error, $\bar{n}_j$ = average number of participants per trial. Only Estimand 1 assessed in scenarios with 30 trials.

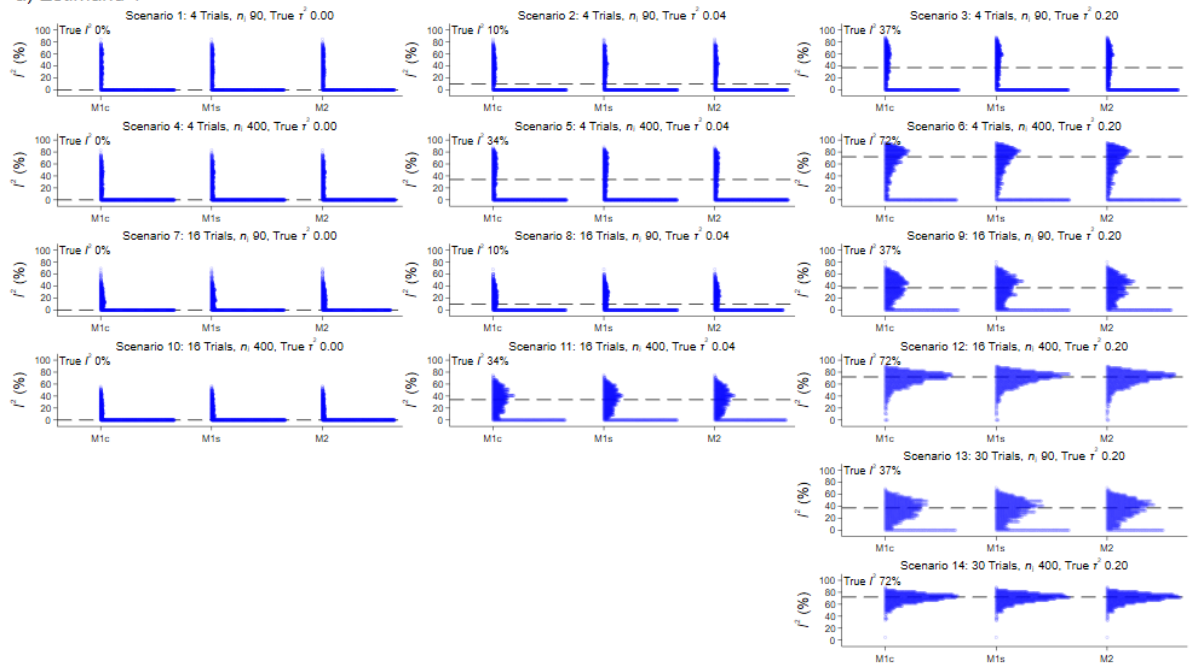

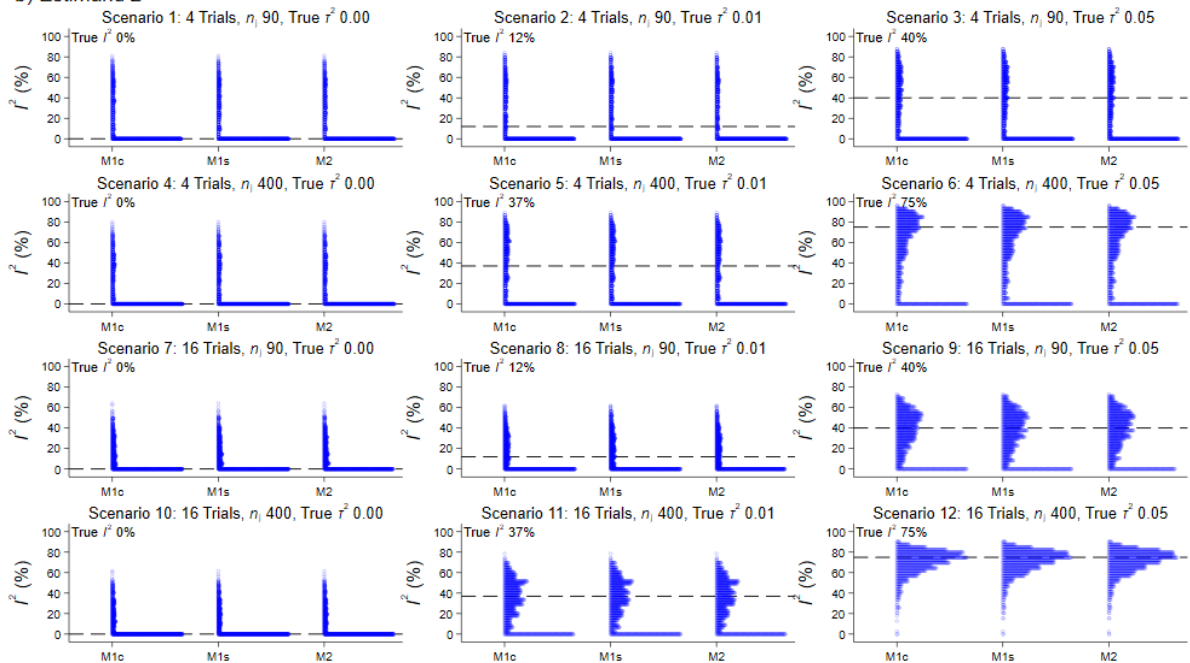

Supplemental Figure 1: Distribution of estimated values of $I^2$ from a) Estimand 1 (treatment-covariate interaction at two years) and b) Estimand 2 (treatment-covariate-time interaction). Each circle represents one replicated dataset. M1c = parsimonious 1-stage model with common residual error variance, M1s = full 1-stage model with trial-level residual error variance, M2 = 2-stage model.

Supplemental Table 4: Bias, empirical standard error, and mean square error for $\tau^2$ (dB).

| Het | $\bar{n}_j$ | Trials | True $\tau^2$ | Parsimonious (M1c) | | | | Full (M1s) | | | | Two-stage (M2) | | | |
|---|---|---|---|---|---|---|---|---|---|---|---|---|---|---|---|
| | | | | Mean | Bias (MCSE) | ESE | MSE | Mean | Bias (MCSE) | ESE | MSE | Mean | Bias (MCSE) | ESE | MSE |
| **Estimand 1** | | | | | | | | | | | | | | | |
| None | 90 | 4 | 0.00 | 0.1151 | 0.1151 (0.0072) | 0.2278 | 0.0651 | 0.1178 | 0.1178 (0.0073) | 0.2304 | 0.0669 | 0.1195 | 0.1195 (0.0073) | 0.2313 | 0.0677 |
| | | 16 | | 0.0529 | 0.0529 (0.0029) | 0.0909 | 0.0111 | 0.0550 | 0.0550 (0.0030) | 0.0941 | 0.0119 | 0.0603 | 0.0603 (0.0031) | 0.0975 | 0.0131 |
| | 400 | 4 | 0.00 | 0.0240 | 0.0240 (0.0014) | 0.0446 | 0.0026 | 0.0239 | 0.0239 (0.0014) | 0.0448 | 0.0026 | 0.0240 | 0.0240 (0.0014) | 0.0448 | 0.0026 |
| | | 16 | | 0.0104 | 0.0104 (0.0006) | 0.0182 | 0.0004 | 0.0104 | 0.0104 (0.0006) | 0.0182 | 0.0004 | 0.0106 | 0.0106 (0.0006) | 0.0184 | 0.0004 |
| Low | 90 | 4 | 0.04 | 0.1370 | 0.0970 (0.0074) | 0.2349 | 0.0645 | 0.1375 | 0.0975 (0.0075) | 0.2373 | 0.0658 | 0.1393 | 0.0993 (0.0075) | 0.2383 | 0.0666 |
| | | 16 | | 0.0733 | 0.0333 (0.0033) | 0.1030 | 0.0117 | 0.0754 | 0.0354 (0.0033) | 0.1057 | 0.0124 | 0.0812 | 0.0412 (0.0034) | 0.1083 | 0.0134 |
| | 400 | 4 | 0.04 | 0.0555 | 0.0155 (0.0027) | 0.0856 | 0.0076 | 0.0557 | 0.0157 (0.0027) | 0.0859 | 0.0076 | 0.0557 | 0.0157 (0.0027) | 0.0859 | 0.0076 |
| | | 16 | | 0.0428 | 0.0028 (0.0013) | 0.0397 | 0.0016 | 0.0427 | 0.0027 (0.0013) | 0.0397 | 0.0016 | 0.0430 | 0.0030 (0.0013) | 0.0397 | 0.0016 |
| High | 90 | 4 | 0.20 | 0.2684 | 0.0684 (0.0122) | 0.3856 | 0.1532 | 0.2677 | 0.0677 (0.0122) | 0.3852 | 0.1528 | 0.2696 | 0.0696 (0.0122) | 0.3855 | 0.1533 |
| | | 16 | | 0.2206 | 0.0206 (0.0060) | 0.1890 | 0.0361 | 0.2200 | 0.0200 (0.0061) | 0.1928 | 0.0375 | 0.2285 | 0.0285 (0.0061) | 0.1925 | 0.0378 |
| | | 30 | | 0.2029 | 0.0029 (0.0045) | 0.1424 | 0.0203 | 0.2031 | 0.0031 (0.0046) | 0.1446 | 0.0209 | 0.2122 | 0.0122 (0.0046) | 0.1442 | 0.0209 |
| | 400 | 4 | 0.20 | 0.1954 | -0.0046 (0.0069) | 0.2198 | 0.0483 | 0.1955 | -0.0045 (0.0070) | 0.2199 | 0.0483 | 0.1955 | -0.0045 (0.0070) | 0.2199 | 0.0483 |
| | | 16 | | 0.2014 | 0.0014 (0.0034) | 0.1070 | 0.0114 | 0.2015 | 0.0015 (0.0034) | 0.1072 | 0.0115 | 0.2017 | 0.0017 (0.0034) | 0.1071 | 0.0115 |
| | | 30 | | 0.1957 | -0.0043 (0.0023) | 0.0721 | 0.0052 | 0.1957 | -0.0043 (0.0023) | 0.0721 | 0.0052 | 0.1959 | -0.0041 (0.0023) | 0.0721 | 0.0052 |
| **Estimand 2** | | | | | | | | | | | | | | | |
| None | 90 | 4 | 0.00 | 0.0240 | 0.0240 (0.0015) | 0.0472 | 0.0028 | 0.0242 | 0.0242 (0.0015) | 0.0473 | 0.0028 | 0.0243 | 0.0243 (0.0015) | 0.0474 | 0.0028 |
| | | 16 | | 0.0116 | 0.0116 (0.0006) | 0.0191 | 0.0005 | 0.0118 | 0.0118 (0.0006) | 0.0192 | 0.0005 | 0.0121 | 0.0121 (0.0006) | 0.0194 | 0.0005 |
| | 400 | 4 | 0.00 | 0.0046 | 0.0046 (0.0003) | 0.0088 | 0.0001 | 0.0046 | 0.0046 (0.0003) | 0.0089 | 0.0001 | 0.0046 | 0.0046 (0.0003) | 0.0089 | 0.0001 |
| | | 16 | | 0.0024 | 0.0024 (0.0001) | 0.0041 | 0.0000 | 0.0024 | 0.0024 (0.0001) | 0.0041 | 0.0000 | 0.0024 | 0.0024 (0.0001) | 0.0041 | 0.0000 |
| Low | 90 | 4 | 0.01 | 0.0318 | 0.0218 (0.0018) | 0.0585 | 0.0039 | 0.0318 | 0.0218 (0.0019) | 0.0587 | 0.0039 | 0.0319 | 0.0219 (0.0019) | 0.0587 | 0.0039 |
| | | 16 | | 0.0172 | 0.0072 (0.0007) | 0.0232 | 0.0006 | 0.0171 | 0.0071 (0.0007) | 0.0234 | 0.0006 | 0.0175 | 0.0075 (0.0007) | 0.0235 | 0.0006 |
| | 400 | 4 | 0.01 | 0.0131 | 0.0031 (0.0006) | 0.0197 | 0.0004 | 0.0131 | 0.0031 (0.0006) | 0.0196 | 0.0004 | 0.0131 | 0.0031 (0.0006) | 0.0196 | 0.0004 |
| | | 16 | | 0.0104 | 0.0004 (0.0003) | 0.0093 | 0.0001 | 0.0104 | 0.0004 (0.0003) | 0.0093 | 0.0001 | 0.0104 | 0.0004 (0.0003) | 0.0093 | 0.0001 |
| High | 90 | 4 | 0.05 | 0.0628 | 0.0128 (0.0028) | 0.0885 | 0.0080 | 0.0626 | 0.0126 (0.0028) | 0.0884 | 0.0080 | 0.0627 | 0.0127 (0.0028) | 0.0884 | 0.0080 |
| | | 16 | | 0.0541 | 0.0041 (0.0014) | 0.0429 | 0.0019 | 0.0539 | 0.0039 (0.0014) | 0.0430 | 0.0019 | 0.0543 | 0.0043 (0.0014) | 0.0430 | 0.0019 |
| | 400 | 4 | 0.05 | 0.0477 | -0.0023 (0.0017) | 0.0534 | 0.0029 | 0.0478 | -0.0022 (0.0017) | 0.0535 | 0.0029 | 0.0478 | -0.0022 (0.0017) | 0.0535 | 0.0029 |
| | | 16 | | 0.0499 | -0.0001 (0.0008) | 0.0256 | 0.0007 | 0.0499 | -0.0001 (0.0008) | 0.0256 | 0.0007 | 0.0499 | -0.0001 (0.0008) | 0.0256 | 0.0007 |

1000 simulated datasets per scenario. Estimand 1 = treatment-covariate interaction at two years, Estimand 2 = treatment-covariate-time interaction. ESE = empirical standard error, Het = between-trial heterogeneity, M1c = parsimonious 1-stage model with common residual error variance, M1s = full 1-stage model with trial-level residual error variance, M2 = 2-stage model, MCSE = Monte Carlo standard error, MSE = mean square error, $\bar{n}_j$ = average number of participants per trial. Only Estimand 1 assessed in scenarios with 30 trials.

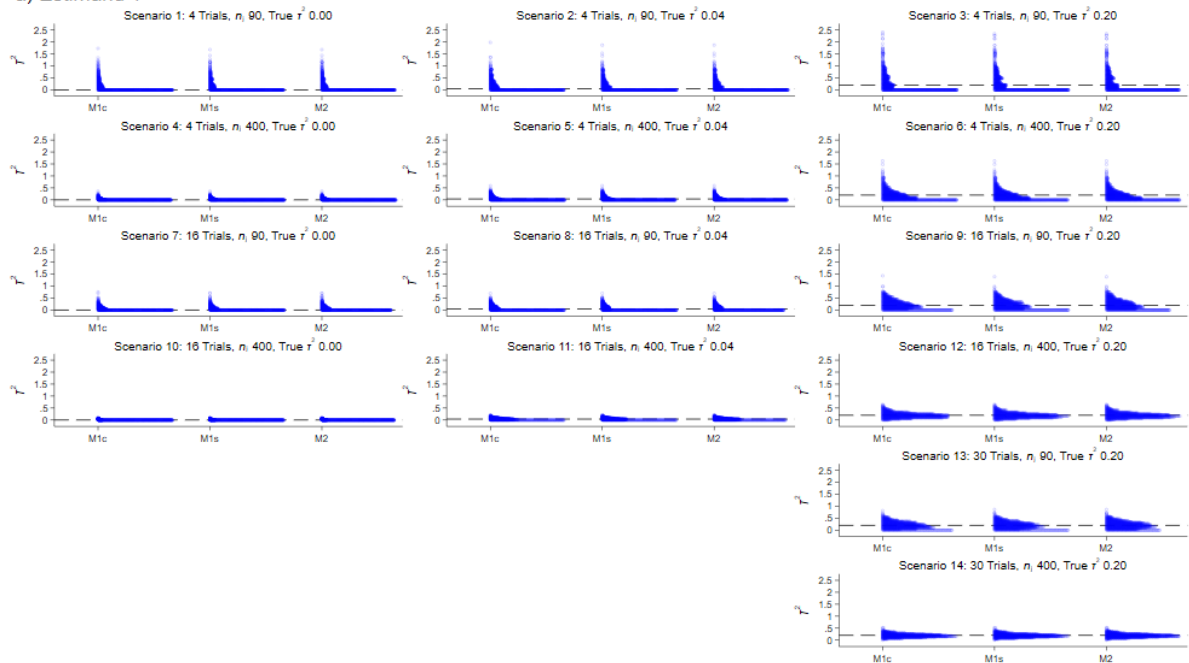

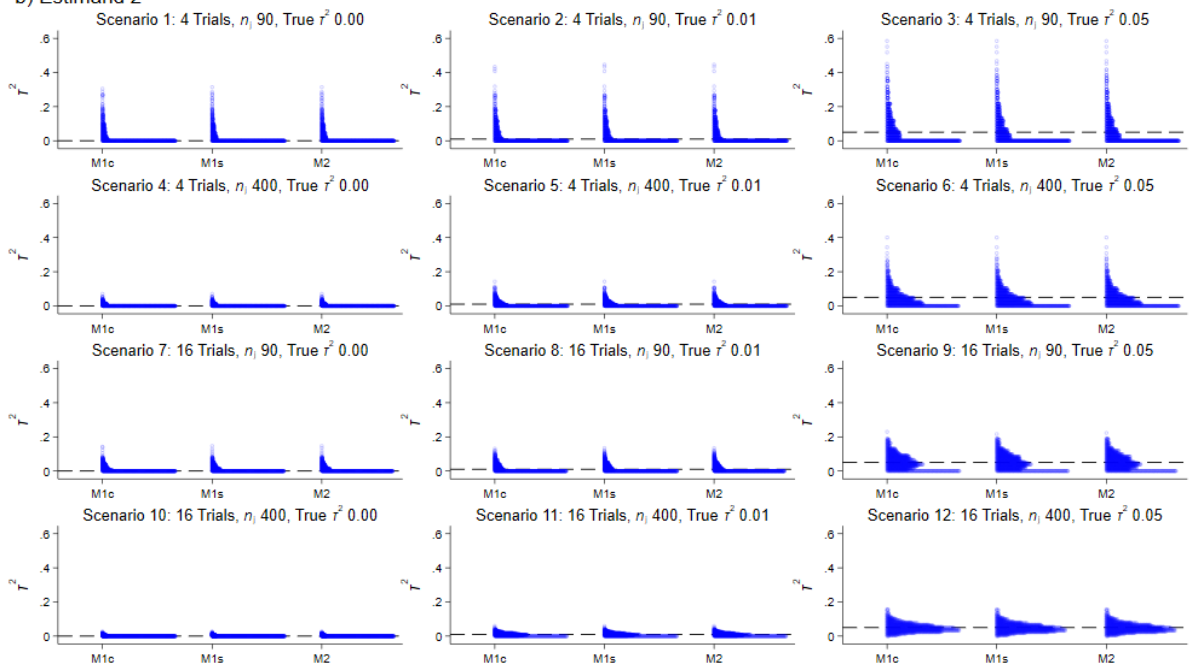

Supplemental Figure 2: Distribution of estimated values of $\tau^2$ (dB) from a) Estimand 1 (treatment-covariate interaction at two years) and b) Estimand 2 (treatment-covariate-time interaction). Each circle represents one replicated dataset. M1c = parsimonious 1-stage model with common residual error variance, M1s = full 1-stage model with trial-level residual error variance, M2 = 2-stage model.

Supplemental Table 5: Bias, empirical standard error, and mean square error for average within-trial variance ($\bar{\sigma}^2_\theta$, dB).

| $\bar{n}_j$ | Trials | Het | True $\bar{\sigma}^2_\theta$ | Parsimonious (M1c) | | | | Full (M1s) | | | | Two-stage (M2) | | | |
|---|---|---|---|---|---|---|---|---|---|---|---|---|---|---|---|
| | | | | Mean | Bias (MCSE) | ESE | MSE | Mean | Bias (MCSE) | ESE | MSE | Mean | Bias (MCSE) | ESE | MSE |
| **Estimand 1** | | | | | | | | | | | | | | | |
| 90 | 4 | None | 0.3413 | 0.3573 | 0.0159 (0.0019) | 0.0590 | 0.0037 | 0.3529 | 0.0116 (0.0018) | 0.0579 | 0.0035 | 0.3584 | 0.0170 (0.0019) | 0.0597 | 0.0039 |
| | | Low | | 0.3592 | 0.0178 (0.0018) | 0.0574 | 0.0036 | 0.3545 | 0.0131 (0.0018) | 0.0565 | 0.0034 | 0.3597 | 0.0184 (0.0018) | 0.0582 | 0.0037 |
| | | High | | 0.3578 | 0.0164 (0.0017) | 0.0552 | 0.0033 | 0.3534 | 0.0120 (0.0017) | 0.0541 | 0.0031 | 0.3585 | 0.0172 (0.0018) | 0.0557 | 0.0034 |
| | 16 | None | 0.3413 | 0.3527 | 0.0114 (0.0008) | 0.0260 | 0.0008 | 0.3455 | 0.0041 (0.0008) | 0.0254 | 0.0007 | 0.3500 | 0.0087 (0.0008) | 0.0262 | 0.0008 |
| | | Low | | 0.3502 | 0.0089 (0.0008) | 0.0243 | 0.0007 | 0.3430 | 0.0017 (0.0007) | 0.0236 | 0.0006 | 0.3474 | 0.0061 (0.0008) | 0.0241 | 0.0006 |
| | | High | | 0.3531 | 0.0118 (0.0008) | 0.0261 | 0.0008 | 0.3456 | 0.0043 (0.0008) | 0.0254 | 0.0007 | 0.3498 | 0.0085 (0.0008) | 0.0260 | 0.0007 |
| | 30 | High | 0.3413 | 0.3506 | 0.0092 (0.0006) | 0.0191 | 0.0004 | 0.3429 | 0.0015 (0.0006) | 0.0184 | 0.0003 | 0.3469 | 0.0056 (0.0006) | 0.0189 | 0.0004 |
| 400 | 4 | None | 0.0768 | 0.0774 | 0.0006 (0.0001) | 0.0036 | 0.0000 | 0.0772 | 0.0004 (0.0001) | 0.0036 | 0.0000 | 0.0775 | 0.0007 (0.0001) | 0.0036 | 0.0000 |
| | | Low | | 0.0772 | 0.0004 (0.0001) | 0.0036 | 0.0000 | 0.0770 | 0.0002 (0.0001) | 0.0036 | 0.0000 | 0.0772 | 0.0004 (0.0001) | 0.0036 | 0.0000 |
| | | High | | 0.0774 | 0.0006 (0.0001) | 0.0039 | 0.0000 | 0.0772 | 0.0004 (0.0001) | 0.0038 | 0.0000 | 0.0774 | 0.0006 (0.0001) | 0.0039 | 0.0000 |
| | 16 | None | 0.0768 | 0.0773 | 0.0005 (0.0001) | 0.0019 | 0.0000 | 0.0769 | 0.0001 (0.0001) | 0.0019 | 0.0000 | 0.0771 | 0.0003 (0.0001) | 0.0019 | 0.0000 |
| | | Low | | 0.0772 | 0.0004 (0.0001) | 0.0018 | 0.0000 | 0.0768 | 0.0000 (0.0001) | 0.0018 | 0.0000 | 0.0770 | 0.0002 (0.0001) | 0.0018 | 0.0000 |
| | | High | | 0.0773 | 0.0005 (0.0001) | 0.0019 | 0.0000 | 0.0769 | 0.0001 (0.0001) | 0.0019 | 0.0000 | 0.0771 | 0.0003 (0.0001) | 0.0019 | 0.0000 |
| | 30 | High | 0.0768 | 0.0772 | 0.0004 (0.0000) | 0.0013 | 0.0000 | 0.0768 | 0.0000 (0.0000) | 0.0013 | 0.0000 | 0.0770 | 0.0002 (0.0000) | 0.0013 | 0.0000 |
| **Estimand 2** | | | | | | | | | | | | | | | |
| 90 | 4 | None | 0.0759 | 0.0796 | 0.0037 (0.0004) | 0.0115 | 0.0001 | 0.0794 | 0.0035 (0.0004) | 0.0114 | 0.0001 | 0.0794 | 0.0036 (0.0004) | 0.0114 | 0.0001 |
| | 4 | Low | | 0.0799 | 0.0041 (0.0004) | 0.0114 | 0.0001 | 0.0797 | 0.0038 (0.0004) | 0.0114 | 0.0001 | 0.0797 | 0.0039 (0.0004) | 0.0114 | 0.0001 |
| | 4 | High | | 0.0797 | 0.0038 (0.0003) | 0.0110 | 0.0001 | 0.0795 | 0.0036 (0.0003) | 0.0109 | 0.0001 | 0.0795 | 0.0036 (0.0003) | 0.0109 | 0.0001 |
| | 16 | None | 0.0759 | 0.0784 | 0.0025 (0.0002) | 0.0051 | 0.0000 | 0.0780 | 0.0022 (0.0002) | 0.0050 | 0.0000 | 0.0780 | 0.0022 (0.0002) | 0.0050 | 0.0000 |
| | 16 | Low | | 0.0779 | 0.0021 (0.0002) | 0.0048 | 0.0000 | 0.0775 | 0.0017 (0.0002) | 0.0048 | 0.0000 | 0.0776 | 0.0017 (0.0002) | 0.0048 | 0.0000 |
| | 16 | High | | 0.0785 | 0.0026 (0.0002) | 0.0052 | 0.0000 | 0.0781 | 0.0022 (0.0002) | 0.0052 | 0.0000 | 0.0781 | 0.0022 (0.0002) | 0.0052 | 0.0000 |
| 400 | 4 | None | 0.0171 | 0.0172 | 0.0001 (0.0000) | 0.0006 | 0.0000 | 0.0172 | 0.0001 (0.0000) | 0.0006 | 0.0000 | 0.0172 | 0.0001 (0.0000) | 0.0006 | 0.0000 |
| | 4 | Low | | 0.0172 | 0.0001 (0.0000) | 0.0006 | 0.0000 | 0.0172 | 0.0001 (0.0000) | 0.0006 | 0.0000 | 0.0172 | 0.0001 (0.0000) | 0.0006 | 0.0000 |
| | 4 | High | | 0.0172 | 0.0001 (0.0000) | 0.0006 | 0.0000 | 0.0172 | 0.0001 (0.0000) | 0.0006 | 0.0000 | 0.0172 | 0.0001 (0.0000) | 0.0006 | 0.0000 |
| | 16 | None | 0.0171 | 0.0172 | 0.0001 (0.0000) | 0.0003 | 0.0000 | 0.0171 | 0.0001 (0.0000) | 0.0003 | 0.0000 | 0.0172 | 0.0001 (0.0000) | 0.0003 | 0.0000 |
| | 16 | Low | | 0.0171 | 0.0001 (0.0000) | 0.0003 | 0.0000 | 0.0171 | 0.0001 (0.0000) | 0.0003 | 0.0000 | 0.0171 | 0.0001 (0.0000) | 0.0003 | 0.0000 |
| | 16 | High | | 0.0172 | 0.0001 (0.0000) | 0.0003 | 0.0000 | 0.0171 | 0.0001 (0.0000) | 0.0003 | 0.0000 | 0.0171 | 0.0001 (0.0000) | 0.0003 | 0.0000 |

1000 simulated datasets per scenario. Estimand 1 = treatment-covariate interaction at two years, Estimand 2 = treatment-covariate-time interaction. ESE = empirical standard error, Het = between-trial heterogeneity, M1c = parsimonious 1-stage model with common residual error variance, M1s = full 1-stage model with trial-level residual error variance, M2 = 2-stage model, MCSE = Monte Carlo standard error, MSE = mean square error, $\bar{n}_j$ = average number of participants per trial. Only Estimand 1 assessed in scenarios with 30 trials.

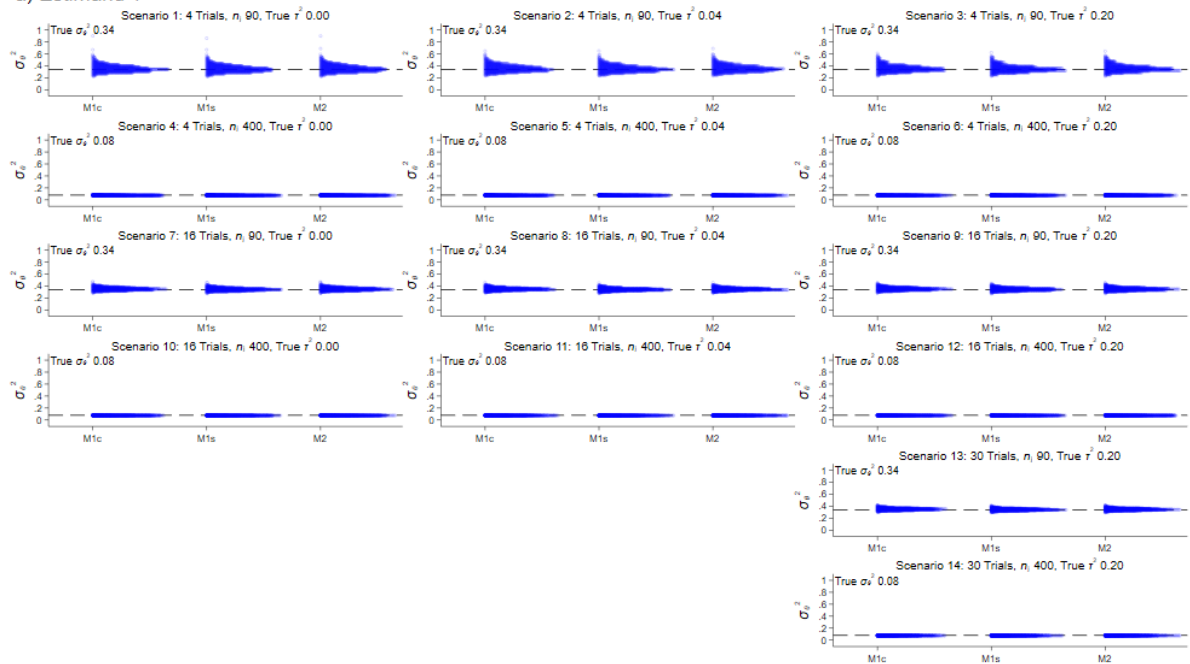

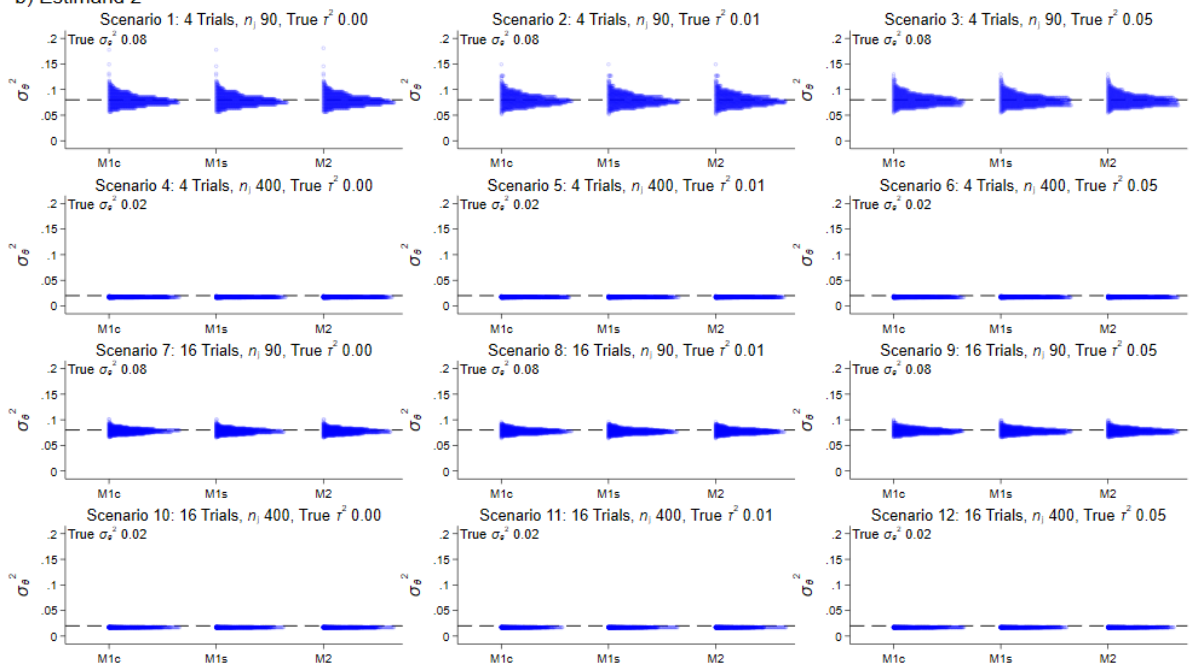

Supplemental Figure 3: Distribution of estimated values of average within-trial variance ($\bar{\sigma}_\theta^2$, dB) from a) Estimand 1 (treatment-covariate interaction at two years) and b) Estimand 2 (treatment-covariate-time interaction). Each circle represents one replicated dataset. M1c = parsimonious 1-stage model with common residual error variance, M1s = full 1-stage model with trial-level residual error variance, M2 = 2-stage model.

Supplemental Table 6: Confidence interval coverage for $\tau^2$ among datasets with non-missing upper and lower 95% confidence limits.

| Scenario | Trials | $\bar{n}_j$ | Het | Parsimonious (M1c) | | Full (M1s) | | Two-stage (M2) | |
|---|---|---|---|---|---|---|---|---|---|
| | | | | n | % (MCSE) | n | % (MCSE) | n | % (MCSE) |
| **Estimand 1** | | | | | | | | | |
| 1 | 4 | 90 | None | 394 | 0.5 (0.4) | 394 | 1.0 (0.5) | 1000 | 98.4 (0.4) |
| 2 | | | Low | 474 | 85.7 (1.6) | 466 | 84.5 (1.7) | 1000 | 98.7 (0.4) |
| 3 | | | High | 601 | 96.5 (0.7) | 604 | 96.5 (0.7) | 1000 | 98.6 (0.4) |
| 4 | | 400 | None | 415 | 0.0 (0.0) | 415 | 0.0 (0.0) | 1000 | 98.7 (0.4) |
| 5 | | | Low | 550 | 95.6 (0.9) | 549 | 95.4 (0.9) | 1000 | 98.4 (0.4) |
| 6 | | | High | 815 | 99.4 (0.3) | 811 | 99.4 (0.3) | 1000 | 98.3 (0.4) |
| 7 | 16 | 90 | None | 479 | 0.0 (0.0) | 492 | 0.0 (0.0) | 1000 | 97.3 (0.5) |
| 8 | | | Low | 571 | 84.2 (1.5) | 562 | 82.9 (1.6) | 1000 | 97.7 (0.5) |
| 9 | | | High | 866 | 95.5 (0.7) | 863 | 94.2 (0.8) | 1000 | 96.8 (0.6) |
| 10 | | 400 | None | 419 | 0.0 (0.0) | 417 | 0.2 (0.2) | 1000 | 97.7 (0.5) |
| 11 | | | Low | 824 | 95.3 (0.7) | 823 | 95.3 (0.7) | 1000 | 97.4 (0.5) |
| 12 | | | High | 996 | 97.4 (0.5) | 997 | 97.6 (0.5) | 1000 | 93.2 (0.8) |
| 13 | 30 | 90 | High | 921 | 95.4 (0.7) | 921 | 95.7 (0.7) | 1000 | 95.0 (0.7) |
| 14 | | 400 | High | 1000 | 97.3 (0.5) | 1000 | 97.3 (0.5) | 1000 | 94.7 (0.7) |
| **Estimand 2** | | | | | | | | | |
| 1 | 4 | 90 | None | 627 | 12.9 (1.3) | 565 | 15.0 (1.5) | 1000 | 98.7 (0.4) |
| 2 | | | Low | 651 | 89.4 (1.2) | 585 | 87.7 (1.4) | 1000 | 98.2 (0.4) |
| 3 | | | High | 777 | 97.9 (0.5) | 716 | 98.0 (0.5) | 1000 | 99.0 (0.3) |
| 4 | | 400 | None | 730 | 6.8 (0.9) | 804 | 8.0 (1.0) | 1000 | 99.4 (0.2) |
| 5 | | | Low | 821 | 97.7 (0.5) | 868 | 97.7 (0.5) | 1000 | 98.9 (0.3) |
| 6 | | | High | 947 | 99.6 (0.2) | 957 | 99.6 (0.2) | 1000 | 98.0 (0.4) |
| 7 | 16 | 90 | None | 881 | 0.1 (0.1) | 846 | 0.1 (0.1) | 1000 | 97.8 (0.5) |
| 8 | | | Low | 899 | 79.5 (1.3) | 882 | 81.3 (1.3) | 1000 | 98.3 (0.4) |
| 9 | | | High | 969 | 93.1 (0.8) | 964 | 93.9 (0.8) | 1000 | 96.5 (0.6) |
| 10 | | 400 | None | 888 | 0.0 (0.0) | 912 | 0.0 (0.0) | 1000 | 98.3 (0.4) |
| 11 | | | Low | 966 | 92.0 (0.9) | 977 | 89.5 (1.0) | 1000 | 96.9 (0.5) |
| 12 | | | High | 1000 | 98.0 (0.4) | 998 | 98.1 (0.4) | 1000 | 93.7 (0.8) |

1000 simulated datasets per scenario; n = number of datasets with non-missing 95% confidence intervals, % = percentage with the true value of $\tau^2$ within the confidence interval. Estimand 1 = treatment-covariate interaction at two years, Estimand 2 = treatment-covariate-time interaction. M1c = parsimonious 1-stage model with common residual error variance, M1s = full 1-stage model with trial-level residual error variance, M2 = 2-stage model, MCSE = Monte Carlo standard error, $\bar{n}_j$ = average number of participants per trial. Only Estimand 1 assessed in scenarios with 30 trials.

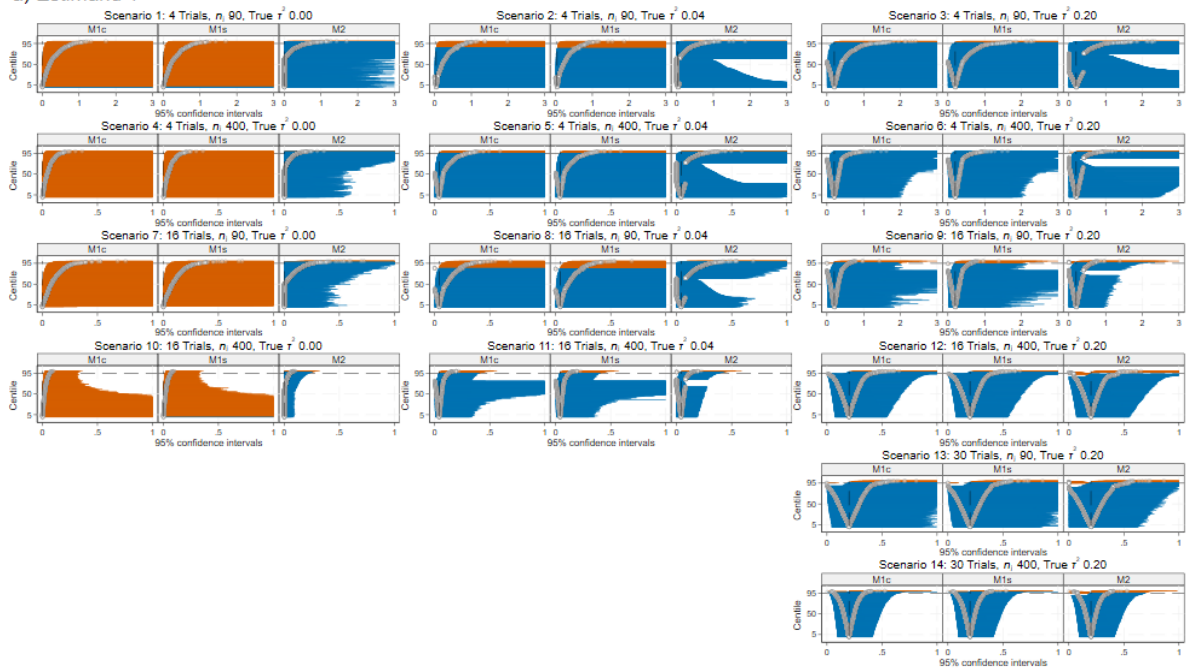

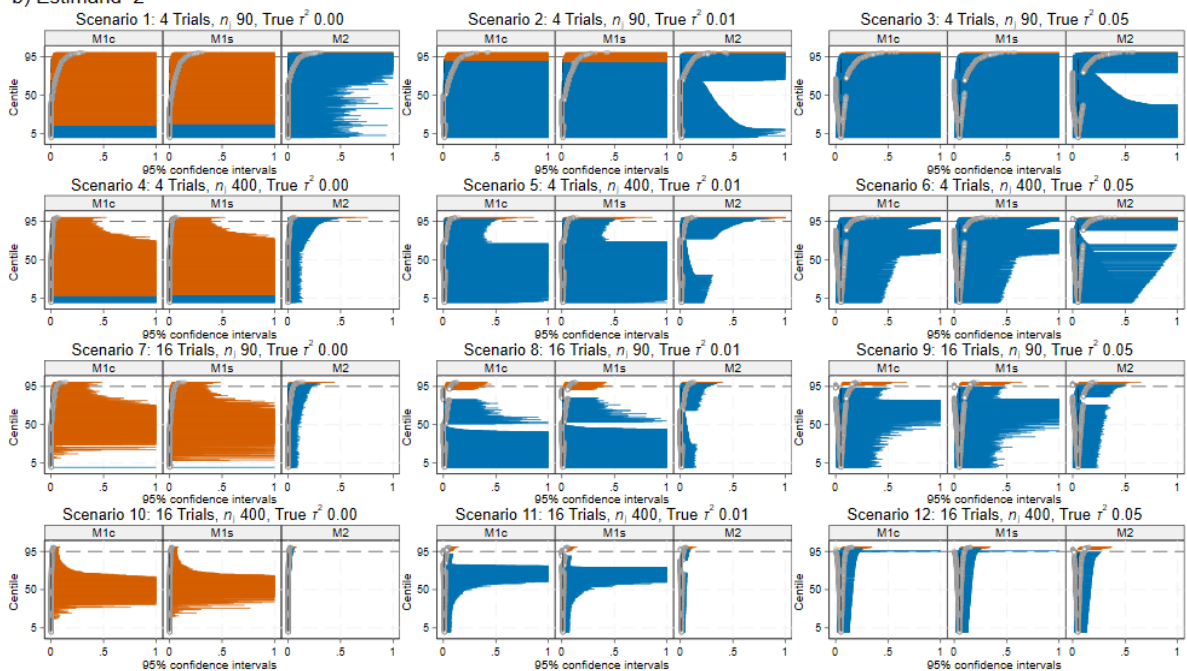

Supplemental Figure 4: 95% confidence interval (CI) zip plots for $\tau^2$ from a) Estimand 1 (treatment-covariate interaction at two years) and b) Estimand 2 (treatment-covariate-time interaction). Only replicated datasets with non-missing confidence limits are shown (≤1000 datasets/scenario). CIs that include the true value of $\tau^2$ (coverers) are indicated in blue, non-coverers are in orange. Grey circles indicate the value of $\tau^2$ estimated in each replicated dataset. Vertical axis is the centile ranked by the magnitude of bias/width of CI. Large upper limits have been truncated. M1c = parsimonious 1-stage model with common residual error variance, M1s = full 1-stage model with trial-level residual error variance, M2 = 2-stage model.

Supplemental Table 7: Agreement in $I^2$ between one- and two-stage models.

| | | | | Difference from 2-stage (percentage points) | | | | | |
|---|---|---|---|---|---|---|---|---|---|
| | | | True | Parsimonious (M1c) | | | Full (M1s) | | |
| Heterogeneity | $\bar{n}_j$ | Trials | $I^2$ (%) | Range | Mean (MCSE) | Median [IQR] | Range | Mean (MCSE) | Median [IQR] |
| **Estimand 1** | | | | | | | | | |
| None | 90 | 4 | 0.0 | -35.7, 43.3 | -0.5 (0.15) | 0.0 [ 0.0, 0.0] | -19.4, 1.0 | -0.1 (0.03) | 0.0 [ 0.0, 0.0] |
| | | 16 | | -43.1, 21.1 | -1.3 (0.15) | 0.0 [-2.4, 0.0] | -9.0, 0.3 | -0.9 (0.04) | 0.0 [-1.3, 0.0] |
| | 400 | 4 | 0.0 | -12.2, 13.2 | 0.0 (0.07) | 0.0 [ 0.0, 0.0] | -1.2, 0.2 | -0.0 (0.00) | 0.0 [-0.0, 0.0] |
| | | 16 | | -9.9, 8.7 | -0.2 (0.06) | 0.0 [-0.2, 0.0] | -2.4, 0.1 | -0.2 (0.01) | 0.0 [-0.3, 0.0] |
| Low | 90 | 4 | 10.5 | -28.1, 20.4 | -0.2 (0.13) | 0.0 [-0.1, 0.0] | -4.1, 1.0 | -0.1 (0.01) | 0.0 [-0.0, 0.0] |
| | | 16 | | -23.9, 15.0 | -1.3 (0.14) | 0.0 [-3.0, 0.0] | -11.0, 0.2 | -0.9 (0.04) | -0.3 [-1.3, 0.0] |
| | 400 | 4 | 34.2 | -15.6, 9.6 | -0.1 (0.06) | 0.0 [-0.2, 0.0] | -1.0, 0.3 | -0.0 (0.00) | 0.0 [ 0.0, 0.0] |
| | | 16 | | -9.3, 13.0 | -0.2 (0.07) | 0.0 [-1.0, 0.7] | -2.6, 0.1 | -0.2 (0.01) | -0.1 [-0.3, 0.0] |
| High | 90 | 4 | 36.9 | -28.8, 28.7 | -0.2 (0.14) | 0.0 [-0.6, 0.4] | -16.0, 1.3 | -0.1 (0.02) | 0.0 [ 0.0, 0.1] |
| | | 16 | | -21.2, 20.6 | -1.0 (0.14) | -0.5 [-3.1, 0.8] | -16.4, 0.4 | -0.9 (0.05) | -0.4 [-1.2, 0.0] |
| | | 30 | | -14.4, 16.3 | -1.3 (0.11) | -1.1 [-3.0, 0.3] | -7.5, 0.3 | -1.0 (0.04) | -0.6 [-1.3,-0.2] |
| | 400 | 4 | 72.3 | -10.7, 9.8 | -0.0 (0.05) | 0.0 [-0.2, 0.2] | -0.8, 0.3 | 0.0 (0.00) | 0.0 [ 0.0, 0.0] |
| | | 16 | | -4.3, 4.5 | -0.1 (0.02) | -0.0 [-0.3, 0.2] | -0.8, 0.1 | 0.0 (0.00) | 0.0 [ 0.0, 0.0] |
| | | 30 | | -2.4, 2.6 | -0.1 (0.01) | -0.1 [-0.3, 0.1] | -0.5, 0.1 | 0.0 (0.00) | 0.0 [ 0.0, 0.0] |
| **Estimand 2** | | | | | | | | | |
| None | 90 | 4 | 0.0 | -17.1, 13.6 | -0.2 (0.06) | 0.0 [ 0.0, 0.0] | -3.1, 0.1 | -0.1 (0.01) | 0.0 [-0.0, 0.0] |
| | | 16 | | -14.7, 11.3 | -0.5 (0.07) | 0.0 [-0.7, 0.0] | -4.5, 0.0 | -0.2 (0.01) | 0.0 [-0.4, 0.0] |
| | 400 | 4 | 0.0 | -5.9, 9.1 | -0.0 (0.04) | 0.0 [ 0.0, 0.0] | -0.9, 0.1 | -0.0 (0.00) | 0.0 [-0.0, 0.0] |
| | | 16 | | -5.9, 6.6 | -0.1 (0.03) | 0.0 [-0.0, 0.0] | -1.1, 0.0 | -0.1 (0.00) | 0.0 [-0.1, 0.0] |
| Low | 90 | 4 | 11.6 | -30.8, 24.9 | -0.0 (0.08) | 0.0 [ 0.0, 0.0] | -1.9, 0.1 | -0.1 (0.01) | 0.0 [-0.0, 0.0] |
| | | 16 | | -10.1, 10.8 | -0.2 (0.07) | 0.0 [-0.8, 0.2] | -3.9, 0.0 | -0.3 (0.01) | -0.1 [-0.4, 0.0] |
| | 400 | 4 | 36.9 | -5.8, 6.7 | -0.0 (0.03) | 0.0 [-0.1, 0.1] | -0.6, 3.5 | -0.0 (0.00) | 0.0 [-0.0, 0.0] |
| | | 16 | | -5.7, 5.4 | -0.1 (0.03) | 0.0 [-0.4, 0.3] | -1.2, 0.0 | -0.1 (0.00) | -0.0 [-0.1,-0.0] |
| High | 90 | 4 | 39.7 | -10.8, 20.3 | 0.0 (0.07) | 0.0 [-0.2, 0.3] | -1.7, 0.1 | -0.1 (0.00) | 0.0 [-0.1, 0.0] |
| | | 16 | | -15.7, 18.2 | -0.2 (0.07) | 0.0 [-1.0, 0.5] | -3.6, 0.0 | -0.2 (0.01) | -0.1 [-0.3,-0.1] |
| | 400 | 4 | 74.6 | -6.6, 7.9 | -0.0 (0.03) | 0.0 [-0.1, 0.1] | -1.5, 0.1 | -0.0 (0.00) | 0.0 [-0.0, 0.0] |
| | | 16 | | -2.0, 2.7 | -0.0 (0.01) | -0.0 [-0.1, 0.1] | -0.8, 0.0 | -0.0 (0.00) | -0.0 [-0.0, 0.0] |

1000 simulated datasets per scenario. Estimand 1 = treatment-covariate interaction at two years, Estimand 2 = treatment-covariate-time interaction. Het = between-trial heterogeneity, M1c = parsimonious 1-stage model with common residual error variance, M1s = full 1-stage model with trial-level residual error variance, MCSE = Monte Carlo standard error, $\bar{n}_j$ = average number of participants per trial. Only Estimand 1 assessed in scenarios with 30 trials. Negative value indicates estimate from 1-stage model was less than that from 2-stage model.

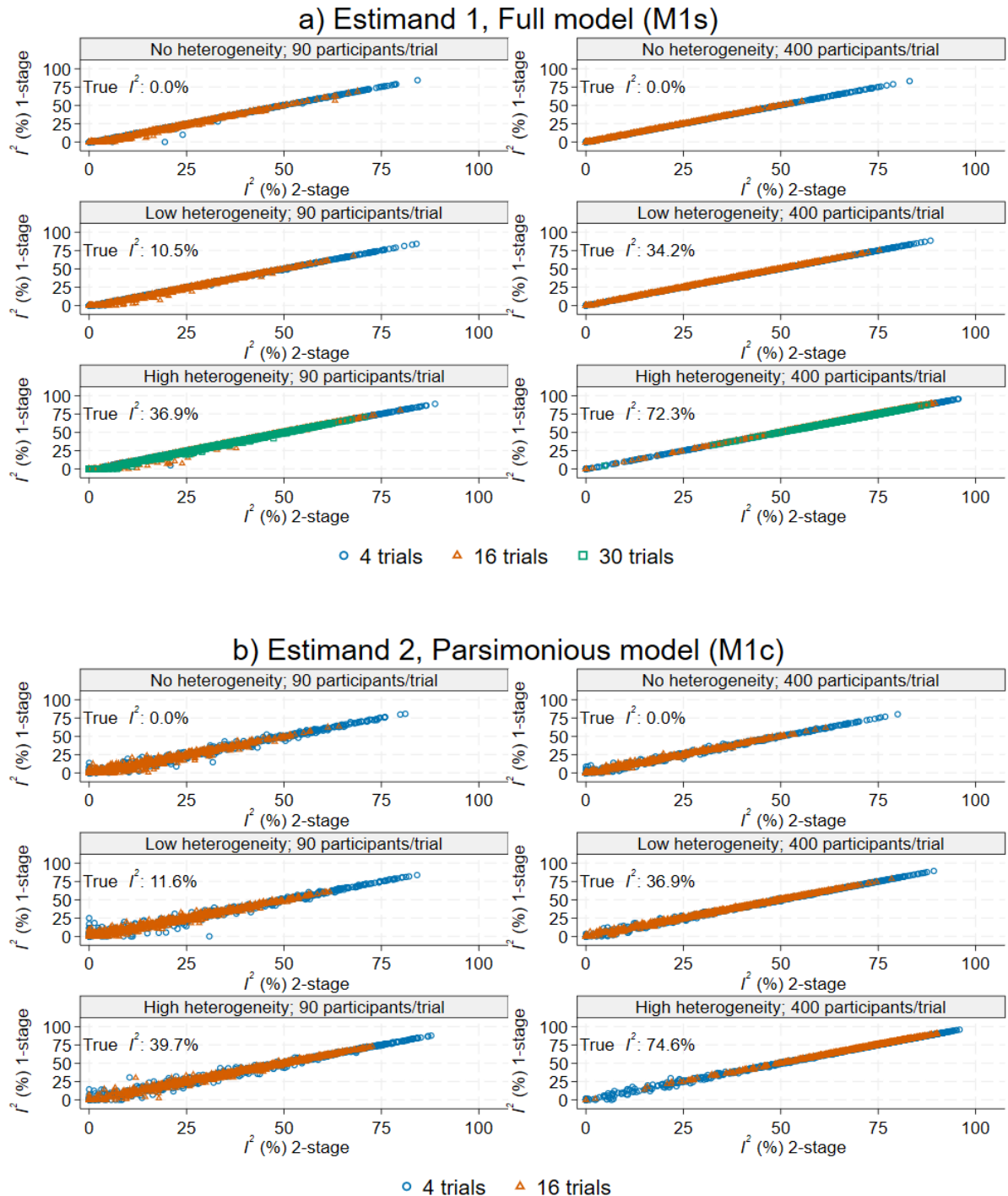

Supplemental Figure 5: Relationship between estimates of $I^2$ for Estimand 1 (treatment-covariate interaction at two years) from the two-stage model and from either a) the full one-stage model with trial-level residual error variance (M1s), or b) the parsimonious one-stage model with common residual error variance (M1c). Each marker represents one simulated dataset.

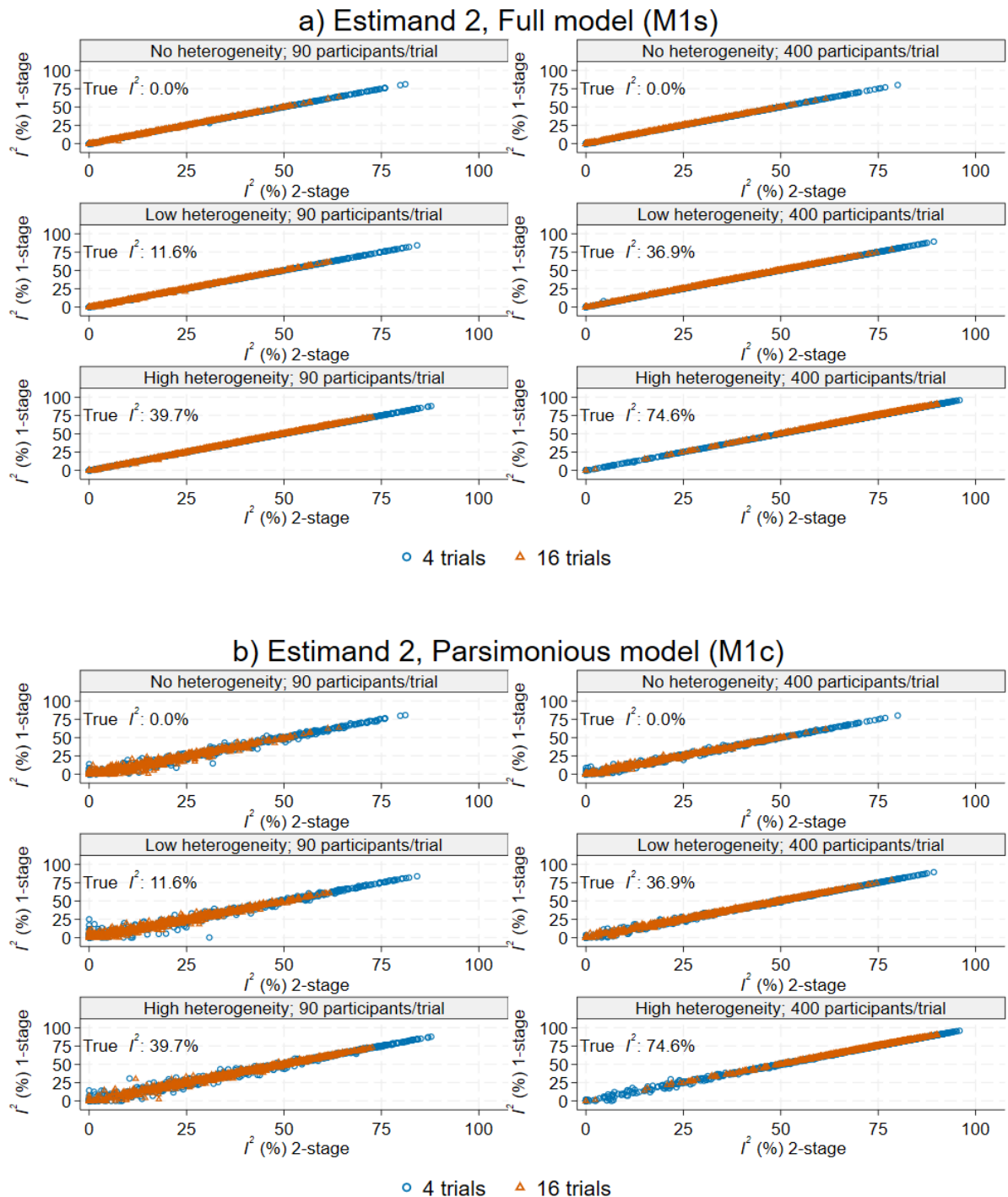

Supplemental Figure 6: Relationship between estimates of $I^2$ for Estimand 2 (treatment-covariate-time interaction) from the two-stage model and from either a) the full one-stage model with trial-level residual error variance (M1s), or b) the parsimonious one-stage model with common residual error variance (M1c). Each marker represents one simulated dataset.